# Optimizing SPARQL Query Answering over OWL Ontologies

**Ilianna Kollia**                                                            ILIANNA2@MAIL.NTUA.GR
*University of Ulm, Germany and*
*National Technical University of Athens, Greece*

**Birte Glimm**                                                              BIRTE.GLIMM@UNI-ULM.DE
*University of Ulm, Germany*

## Abstract

The SPARQL query language is currently being extended by the World Wide Web Consortium (W3C) with so-called entailment regimes. An entailment regime defines how queries are evaluated under more expressive semantics than SPARQL's standard simple entailment, which is based on subgraph matching. The queries are very expressive since variables can occur within complex concepts and can also bind to concept or role names.

In this paper, we describe a sound and complete algorithm for the OWL Direct Semantics entailment regime. We further propose several novel optimizations such as strategies for determining a good query execution order, query rewriting techniques, and show how specialized OWL reasoning tasks and the concept and role hierarchy can be used to reduce the query execution time. For determining a good execution order, we propose a cost-based model, where the costs are based on information about the instances of concepts and roles that are extracted from a model abstraction built by an OWL reasoner. We present two ordering strategies: a static and a dynamic one. For the dynamic case, we improve the performance by exploiting an individual clustering approach that allows for computing the cost functions based on one individual sample from a cluster.

We provide a prototypical implementation and evaluate the efficiency of the proposed optimizations. Our experimental study shows that the static ordering usually outperforms the dynamic one when accurate statistics are available. This changes, however, when the statistics are less accurate, e.g., due to nondeterministic reasoning decisions. For queries that go beyond conjunctive instance queries we observe an improvement of up to three orders of magnitude due to the proposed optimizations.

## 1. Introduction

Query answering is important in the context of the Semantic Web since it provides a mechanism via which users and applications can interact with ontologies and data. Several query languages have been designed for this purpose, including RDQL (Seaborne, 2004), SeRQL (Broekstra & Kampman, 2006) and, most recently, SPARQL. In this paper, we consider the SPARQL query language (Prud'hommeaux & Seaborne, 2008), which was standardized in 2008 by the World Wide Web Consortium (W3C) and which is currently being extended to SPARQL 1.1 (Harris & Seaborne, 2013). Since 2008, SPARQL has developed into the main query language for the Semantic Web and is now supported by most RDF triple stores. The query evaluation mechanism defined in the SPARQL Query specification is based on subgraph matching. This form of query evaluation is also called simple entailment since it can equally be defined in terms of the simple entailment relation between RDF graphs (Hayes, 2004). SPARQL 1.1 includes several *entailment regimes* (Glimm & Ogbuji, 2013) in order to use more elaborate entailment relations, such as those induced by RDF Schema (RDFS) (Brickley & Guha, 2004) or OWL (Motik, Patel-Schneider, & Cuenca Grau, 2012b; Schneider, 2012). Query answering under such entailment regimes is more complex as it may involve retrieving answers that only follow implicitly from the queried graph, which is seen as an OWL ontology when using OWL entailment. While several implementations for SPARQL's RDFS entailment regime are available (e.g., Oracle 11g (Oracle, 2013), Apache Jena (The Apache Software Foundation, 2013), or Stardog (Clark & Parsia, 2013b)), the development of tools that provide full SPARQL support under OWL semantics is still an ongoing effort.

Since we consider the OWL Direct Semantics entailment regime of SPARQL 1.1 in this paper, when we talk about SPARQL queries or the evaluation of SPARQL queries, we always assume that the OWL Direct Semantics entailment regime is used. In this setting, the WHERE clause of a query can be seen as a set of extended OWL axioms (an extended OWL ontology), which can have variables in place of concept, role or individual names. The query answers contain each instantiation of the variables that leads to OWL axioms that are entailed by the queried ontology. Thus, a naive query evaluation procedure can be realized through OWL's standard reasoning task of entailment checking.

Please note that there are two types of individual variables in SPARQL; standard (distinguished) variables and anonymous individuals (aka blank nodes). The anonymous individuals are treated like distinguished variables with the difference that they cannot be selected and, hence, their bindings cannot appear in the query answer. This is in contrast to conjunctive queries, where anonymous individuals are treated as existential variables. On the other hand, anonymous individuals can occur in the query answer as bindings to distinguished variables, i.e., SPARQL treats anonymous individuals from the queried ontology as constants. This treatment of anonymous individuals has been chosen for compatibility with SPARQL's standard subgraph matching semantics. For example, in order to implement the RDF(S) entailment regime, systems can simply extend the queried graph with inferred information (materialization) and can then use SPARQL's standard evaluation mechanism over the materialized graph in order to compute the query results. Similarly, when users move on to systems that support the OWL RL profile (Motik, Cuenca Grau, Horrocks, Wu, Fokoue, & Lutz, 2012a), the OWL RL rule set from the OWL 2 specification can be used to compute the query answers (again via materialization). If one were to change the semantics of blank nodes for SPARQL's entailment regimes to reflect conjunctive query semantics, one could no longer use materialization plus a standard SPARQL query processor to implement the entailment regime. If one were to change the semantics of blank nodes only for the OWL Direct Semantics entailment regime, where materialization cannot be used to implement the regime, users would not simply get more answers by moving from systems that support RDF(S) to systems that support OWL's Direct Semantics, but it could also happen that they get less answers by using a more expressive logic, which is counter-intuitive.

Over the last decade, much effort has been spent on optimizing standard reasoning tasks such as entailment checking, classification, or realization (i.e., the computation of instances of all concepts and roles) (Sirin, Cuenca Grau, & Parsia, 2006; Tsarkov, Horrocks, & Patel-Schneider, 2007; Glimm, Horrocks, Motik, Shearer, & Stoilos, 2012). The optimization of query answering algorithms has, however, mostly been addressed for conjunctive queries in OWL profiles, most notably the OWL 2 QL profile (Calvanese, Giacomo, Lembo, Lenzerini,

& Rosati, 2007; Kontchakov, Lutz, Toman, Wolter, & Zakharyaschev, 2010; Pérez-Urbina, Motik, & Horrocks, 2010; Rodriguez-Muro & Calvanese, 2012). An exception to this are the works on nRQL and SPARQL-DL. The query language nRQL is supported by Racer Pro (Haarslev, Möller, & Wessel, 2004) and SPARQL-DL is implemented in the Pellet reasoner (Sirin, Parsia, Grau, Kalyanpur, & Katz, 2007). We discuss this in greater detail in Section 8.

In this paper, we address the problem of efficient SPARQL query evaluation for OWL 2 DL ontologies by proposing a range of novel optimizations that deal in particular with the expressive features of SPARQL such as variables in place of concepts or roles. We further adapt common techniques from databases such as cost-based query planning. The costs for our cost model are based on information about the instances of concepts and roles that are extracted from a model abstraction built by an OWL reasoner. We present a static and a dynamic algorithm for finding an optimal or near optimal execution order and for the dynamic case, we improve the performance by exploiting an individual clustering approach that allows for computing the cost functions based on one individual sample from a cluster. We further propose query rewriting techniques and show how specialized OWL reasoning tasks and the concept and role hierarchy can be used to reduce the query execution time. We provide a prototypical implementation and evaluate the efficiency of the proposed optimizations. Our experimental study shows that the static ordering usually outperforms the dynamic one when accurate statistics are available. This changes, however, when the statistics are less accurate, e.g., due to non-deterministic reasoning decisions. For queries that go beyond conjunctive SPARQL instance queries, we observe an improvement of up to three orders of magnitude due to the proposed optimizations.

Note that this paper combines and extends two conference papers: I. Kollia and B. Glimm: Cost based Query Ordering over OWL Ontologies. Proceedings of the 11th International Semantic Web Conference, 2012 and I. Kollia, B. Glimm and I. Horrocks: SPARQL Query Answering over OWL Ontologies. Proceedings of the 8th Extended Semantic Web Conference, 2011. In the current paper we have, additionally to the first above mentioned paper, defined cost functions for general SPARQL queries (i.e., not only for conjunctive instance queries) and added experimental results for these expressive queries. In comparison to the second of the above mentioned papers, we have defined the notion of concept and role polarity and presented theorems that let us prune the search space of possible mappings for axiom templates based on the polarity together with an algorithm that shows the way we use the optimization. Moreover, more experimental results have been added for complex queries that make use of this optimization.

The remainder of the paper is organized as follows: we next present some preliminaries, we then present a general query evaluation algorithm in Section 3 that serves as the basis for further optimization. In Section 4, we present the foundations for our cost model, which we then specify in Section 5. In Section 6, we present optimizations for complex queries that cannot directly be mapped to specialized reasoner tasks. Finally, we evaluate our approach in Section 7 and discuss related work in Section 8 before we conclude in Section 9.

## 2. Preliminaries

In this section, we first give a brief introduction into Description Logics since the OWL Direct Semantics is based on the Description Logic $\mathcal{SROIQ}$ (Horrocks, Kutz, & Sattler, 2006). The optimizations we present do not need all features of $\mathcal{SROIQ}$. Hence, we only present $\mathcal{SHOIQ}$, which allows for a shorter and easier to follow presentation.

After introducing $\mathcal{SHOIQ}$, we clarify the relationship between RDF, SPARQL and OWL, we present SPARQL's OWL Direct Semantics entailment regime and we give an overview of the model building tableau and hypertableau calculi.

### 2.1 The Description Logic $\mathcal{SHOIQ}$

We first define the syntax and semantics of roles, and then go on to $\mathcal{SHOIQ}$-concepts, individuals, and ontologies/knowledge bases.

**Definition 1 (Syntax of $\mathcal{SHOIQ}$ ).** *Let $N_C$, $N_R$, and $N_I$ be countable, infinite, and pairwise disjoint sets of* concept names, role names, *and* individual names, *respectively. We call $\mathcal{S} = (N_C, N_R, N_I)$ a* signature. *The set $\mathsf{rol}(\mathcal{S})$ of $\mathcal{SHOIQ}$-roles over $\mathcal{S}$ (or roles for short) is $N_R \cup \{r^- \mid r \in N_R\} \cup \{\top_r, \bot_r\}$, where roles of the form $r^-$ are called* inverse *roles, $\top_r$ is the top role (analogous to owl:topObjectProperty), and $\bot_r$ is the bottom role (analogous to owl:bottomObjectProperty). A role inclusion axiom is of the form $r \sqsubseteq s$ with $r, s$ roles. A transitivity axiom is of the form $\mathsf{trans}(r)$ for $r$ a role. A role hierarchy $\mathcal{H}$ is a finite set of role inclusion and transitivity axioms.*

*For a role hierarchy $\mathcal{H}$, we define the function $\mathsf{inv}$ over roles as $\mathsf{inv}(r) := r^-$ if $r \in N_R$ and $\mathsf{inv}(r) := s$ if $r = s^-$ for a role name $s \in N_R$. Further, we define $\sqsubseteq_{\mathcal{H}}$ as the smallest transitive reflexive relation on roles such that $r \sqsubseteq s \in \mathcal{H}$ implies $r \sqsubseteq_{\mathcal{H}} s$ and $\mathsf{inv}(r) \sqsubseteq_{\mathcal{H}} \mathsf{inv}(s)$. We write $r \equiv_{\mathcal{H}} s$ if $r \sqsubseteq_{\mathcal{H}} s$ and $s \sqsubseteq_{\mathcal{H}} r$. A role $r$ is transitive w.r.t. $\mathcal{H}$ (notation $r^+ \sqsubseteq_{\mathcal{H}} r$) if a role $s$ exists such that $r \sqsubseteq_{\mathcal{H}} s$, $s \sqsubseteq_{\mathcal{H}} r$, and $\mathsf{trans}(s) \in \mathcal{H}$ or $\mathsf{trans}(\mathsf{inv}(s)) \in \mathcal{H}$. A role $s$ is called* simple *w.r.t. $\mathcal{H}$ if there is no role $r$ such that $r$ is transitive w.r.t. $\mathcal{H}$ and $r \sqsubseteq_{\mathcal{H}} s$.*

*Given a signature $\mathcal{S} = (N_C, N_R, N_I)$ and a role hierarchy $\mathcal{H}$, the set of $\mathcal{SHOIQ}$-concepts (or concepts for short) over $\mathcal{S}$ is the smallest set built inductively over symbols from $\mathcal{S}$ using the following grammar, where $o \in N_I, A \in N_C, n \in \mathbb{N}_0$, $s$ is a simple role w.r.t. $\mathcal{H}$, and $r$ is a role w.r.t. $\mathcal{H}$:*

$$C ::= \quad \top \mid \bot \mid \{o\} \mid A \mid \neg C \mid C \sqcap C \mid C \sqcup C \mid \forall r.C \mid \exists r.C \mid \leqslant n\, s.C \mid \geqslant n\, s.C.$$

We now define the semantics of $\mathcal{SHOIQ}$ concepts:

**Definition 2 (Semantics of $\mathcal{SHOIQ}$-concepts).** *An* interpretation $\mathcal{I} = (\Delta^{\mathcal{I}}, \cdot^{\mathcal{I}})$ *consists of a non-empty set $\Delta^{\mathcal{I}}$, the* domain *of $\mathcal{I}$, and a function $\cdot^{\mathcal{I}}$, which maps every concept name $A \in N_C$ to a subset $A^{\mathcal{I}} \subseteq \Delta^{\mathcal{I}}$, every role name $r \in N_R$ to a binary relation $r^{\mathcal{I}} \subseteq \Delta^{\mathcal{I}} \times \Delta^{\mathcal{I}}$, and every individual name $a \in N_I$ to an element $a^{\mathcal{I}} \in \Delta^{\mathcal{I}}$. The top role $\top_r$ is interpreted as $\{\langle \delta, \delta' \rangle \mid \delta, \delta' \in \Delta^{\mathcal{I}}\}$ and the bottom role $\bot_r$ as $\emptyset$. For each role name $r \in N_R$, the interpretation of its inverse role $(r^-)^{\mathcal{I}}$ consists of all pairs $\langle \delta, \delta' \rangle \in \Delta^{\mathcal{I}} \times \Delta^{\mathcal{I}}$ for which $\langle \delta', \delta \rangle \in r^{\mathcal{I}}$.*

*The semantics of $\mathcal{SHOIQ}$-concepts over a signature $\mathcal{S}$ is defined as follows:*

$$\top^{\mathcal{I}} = \Delta^{\mathcal{I}} \qquad\qquad \bot^{\mathcal{I}} = \emptyset \qquad\qquad (\{o\})^{\mathcal{I}} = \{o^{\mathcal{I}}\}$$
$$(\neg C)^{\mathcal{I}} = \Delta^{\mathcal{I}} \setminus C^{\mathcal{I}} \qquad (C \sqcap D)^{\mathcal{I}} = C^{\mathcal{I}} \cap D^{\mathcal{I}} \qquad (C \sqcup D)^{\mathcal{I}} = C^{\mathcal{I}} \cup D^{\mathcal{I}}$$
$$(\forall r.C)^{\mathcal{I}} = \{\delta \in \Delta^{\mathcal{I}} \mid \text{if } \langle \delta, \delta' \rangle \in r^{\mathcal{I}}, \text{ then } \delta' \in C^{\mathcal{I}}\}$$
$$(\exists r.C)^{\mathcal{I}} = \{\delta \in \Delta^{\mathcal{I}} \mid \text{there is a } \langle \delta, \delta' \rangle \in r^{\mathcal{I}} \text{ with } \delta' \in C^{\mathcal{I}}\}$$
$$(\leqslant n\ s.C)^{\mathcal{I}} = \{\delta \in \Delta^{\mathcal{I}} \mid \sharp(s^{\mathcal{I}}(\delta, C)) \leq n\}$$
$$(\geqslant n\ s.C)^{\mathcal{I}} = \{\delta \in \Delta^{\mathcal{I}} \mid \sharp(s^{\mathcal{I}}(\delta, C)) \geq n\}$$

*where $\sharp(M)$ denotes the cardinality of the set $M$ and $s^{\mathcal{I}}(\delta, C)$ is defined as*

$$\{\delta' \in \Delta^{\mathcal{I}} \mid \langle \delta, \delta' \rangle \in s^{\mathcal{I}} \text{ and } \delta' \in C^{\mathcal{I}}\}.$$

**Definition 3** (**Syntax and Semantics of Axioms and Ontologies, Entailment**)**.** *For $C, D$ concepts, a* (general) concept inclusion axiom *(GCI) is an expression $C \sqsubseteq D$. We introduce $C \equiv D$ as an abbreviation for $C \sqsubseteq D$ and $D \sqsubseteq C$. A finite set of GCIs is called a* TBox*. An (ABox) (concept or role) assertion axiom *is an expression of the form $C(a)$, $r(a,b)$, $\neg r(a,b)$, $a \approx b$, or $a \not\approx b$, where $C \in N_C$ is a concept, $r \in N_R$ is a role, and $a, b \in N_I$ are individual names. An* ABox *is a finite set of assertion axioms. An* ontology $\mathcal{O}$ *is a triple $(\mathcal{T}, \mathcal{H}, \mathcal{A})$ with $\mathcal{T}$ a TBox, $\mathcal{H}$ a role hierarchy, and $\mathcal{A}$ an ABox. We use $N_C^{\mathcal{O}}$, $N_R^{\mathcal{O}}$, and $N_I^{\mathcal{O}}$ to denote, respectively, the set of concept, role, and individual names occurring in $\mathcal{O}$.*

*Let $\mathcal{I} = (\Delta^{\mathcal{I}}, \cdot^{\mathcal{I}})$ be an interpretation. Then $\mathcal{I}$ satisfies a* role inclusion axiom $r \sqsubseteq s$ *if $r^{\mathcal{I}} \subseteq s^{\mathcal{I}}$, $\mathcal{I}$ satisfies a* transitivity axiom $\mathsf{trans}(r)$ *if $r^{\mathcal{I}}$ is a transitive binary relation, and a* role hierarchy $\mathcal{H}$ *if it satisfies all role inclusion and transitivity axioms in $\mathcal{H}$. The interpretation $\mathcal{I}$ satisfies a* GCI $C \sqsubseteq D$ *if $C^{\mathcal{I}} \subseteq D^{\mathcal{I}}$; and $\mathcal{I}$ satisfies a* TBox $\mathcal{T}$ *if it satisfies each GCI in $\mathcal{T}$. The interpretation $\mathcal{I}$ satisfies an* assertion axiom $C(a)$ *if $a^{\mathcal{I}} \in C^{\mathcal{I}}$, $r(a,b)$ if $\langle a^{\mathcal{I}}, b^{\mathcal{I}} \rangle \in r^{\mathcal{I}}$, $\neg r(a,b)$ if $\langle a^{\mathcal{I}}, b^{\mathcal{I}} \rangle \notin r^{\mathcal{I}}$, $a \approx b$ if $a^{\mathcal{I}} = b^{\mathcal{I}}$, and $a \not\approx b$ if $a^{\mathcal{I}} \neq b^{\mathcal{I}}$; $\mathcal{I}$ satisfies an* ABox *if it satisfies each assertion in $\mathcal{A}$. We say that $\mathcal{I}$ satisfies $\mathcal{O}$ if $\mathcal{I}$ satisfies $\mathcal{T}$, $\mathcal{H}$, and $\mathcal{A}$. In this case, we say that $\mathcal{I}$ is a* model *of $\mathcal{O}$ and write $\mathcal{I} \models \mathcal{O}$. We say that $\mathcal{O}$ is* consistent *if $\mathcal{O}$ has a model.*

*Given an axiom $\alpha$, we say that $\mathcal{O}$* entails $\alpha$ *(written $\mathcal{O} \models \alpha$) if every model $\mathcal{I}$ of $\mathcal{O}$ satisfies $\alpha$.*

Description Logics can further be extended with concrete domains, which correspond to OWL's datatypes. In such a case, one distinguishes between *abstract roles* that relate two individuals and *concrete roles* that relate an individual with a data value. The Description Logic $\mathcal{SROIQ}$ further allows for a number of features such as role chains of the form hasFather ∘ hasBrother $\sqsubseteq$ hasUncle, support for the special concept Self, which can be used in axioms of the form Narcissist $\sqsubseteq$ $\exists$loves.Self, or for defining roles that are reflexive, irreflexive, symmetric, or asymmetric.

Description Logic ontologies can equally be expressed in terms of OWL ontologies, which in turn can be mapped into RDF graphs (Patel-Schneider & Motik, 2012). The other direction is, however, not always possible, i.e., a mapping from RDF graphs to OWL ontologies is only defined for certain well-formed RDF graphs that correspond to an OWL 2 DL ontology.

## 2.2 The Relationship between RDF, SPARQL, and OWL

SPARQL queries are evaluated over RDF graphs which remain the basic data structure even when adopting a more elaborate semantic interpretation.

**Definition 4 (RDF Graphs).** *RDF is based on the set $I$ of* International Resource Identifiers *(IRIs), the set $L$ of* RDF literals, *and the set $B$ of* blank nodes. *The set $T$ of* RDF terms *is $I \cup L \cup B$. An* RDF graph *is a set of* RDF triples *of the form* (*subject, predicate, object*) $\in (I \cup B) \times I \times T$.

We generally abbreviate IRIs using prefixes rdf, rdfs, owl, and xsd to refer to the RDF, RDFS, OWL, and XML Schema Datatypes namespaces, respectively. The empty prefix is used for an imaginary example namespace, which we completely omit in Description Logic syntax.

An example of a SPARQL query is

SELECT ?x FROM <ontologyIRI> WHERE { ?x rdf:type :C . ?x :r ?y }

The WHERE clause of the SPARQL query consists of a basic graph pattern (BGP): an RDF graph written in Turtle syntax (Beckett, Berners-Lee, Prud'hommeaux, & Carothers, 2013), where some nodes or edges are replaced by variables. A basic graph pattern is more precisely defined as follows:

**Definition 5 (Basic Graph Pattern).** *Let $V$ be a countably infinite set of* query variables *disjoint from $T$. A* triple pattern *is a member of the set $(T \cup V) \times (I \cup V) \times (T \cup V)$, and a* basic graph pattern *(BGP) is a set of triple patterns.*

We do not recall the complete surface syntax of SPARQL here since the only part that is specific to the evaluation of SPARQL queries under OWL's Direct Semantics is the evaluation of BGPs. More complex WHERE clauses, which use operators such as UNION for alternative selection criteria or OPTIONAL to query for optional bindings (Prud'hommeaux & Seaborne, 2008), can be evaluated simply by combining the results obtained by the BGP evaluation. Similarly, operations such as the projection of variables from the SELECT clause is a straightforward operation over the results of the evaluation of the WHERE clause. Therefore, we focus here on BGP evaluation only. For a more detailed introduction to SPARQL queries and their algebra we refer interested readers to the work of Hitzler, Krötzsch, and Rudolph (2009) or Glimm and Krötzsch (2010).

Since the Direct Semantics of OWL is defined in terms of OWL structural objects, i.e., OWL axioms, we map the BGPs of SPARQL queries into structural objects, which can have variables in place of class (concept), object or data property (abstract or concrete role), or individual names or literals. Since there is a direct mapping between OWL axioms and Description Logic axioms, BGPs can be expressed as Description Logic axioms in which variables can occur in place of concept, role and individual names. For example, the BGP of the previous example is mapped to ClassAssertion($C$ ?x) and ObjectPropertyAssertion($r$ ?x ?y) in functional-style syntax or to $C(?x)$ and $r(?x, ?y)$ in Description Logic syntax.

For further details, we refer interested readers to the W3C specification that defines the mapping between OWL structural objects and RDF graphs (Patel-Schneider & Motik, 2012) and to the specification of the OWL Direct Semantics entailment regime of SPARQL (Glimm & Ogbuji, 2013) that defines the extension of this mapping between BGPs and OWL objects with variables.

## 2.3 SPARQL Queries

In the following, we directly write BGPs in Description Logic notation extended to allow for variables in place of concept, role and individual names in axioms. It is worth reminding that SPARQL does not support existentially quantified variables, which is in contrast to database-style conjunctive queries, where one typically also has existential/non-distinguished variables.

For brevity and without loss of generality, we assume here that neither the query nor the queried ontology contains anonymous individuals. We further do not consider data properties and literals, but the presented optimizations can easily be transferred to this case.

**Definition 6 (Query).** *Let $\mathcal{S} = (N_C, N_R, N_I)$ be a signature. A query signature $\mathcal{S}_q$ w.r.t. $\mathcal{S}$ is a six-tuple $(N_C, N_R, N_I, V_C, V_R, V_I)$, where $V_C$, $V_R$, and $V_I$ are countable, infinite, and pairwise disjoint sets of concept variables, role variables, and individual variables disjoint from $N_C$, $N_R$, and $N_I$. A concept term is an element from $N_C \cup V_C$. A role term is an element from $N_R \cup V_R$. An individual term is an element from $N_I \cup V_I$. An axiom template over $\mathcal{S}_q$ is a $\mathcal{SROIQ}$ axiom over $\mathcal{S}$, where one can also use concept variables from $V_C$ in place of concept names, role variables from $V_R$ in place of role names, and individual variables from $V_I$ in place of individual names. A query $q$ w.r.t. a query signature $\mathcal{S}_q$ is a non-empty set of axiom templates over $\mathcal{S}_q$. We use $Vars(q)$ ($Vars(at)$ for an axiom template at) to denote the set of all variables in $q$ (at) and $|q|$ to denote the number of axiom templates in $q$. Let $t, t'$ be individual terms; we call axiom templates of the form $A(t)$ with $A \in N_C$, $r(t, t')$ with $r \in N_R$, or $t \approx t'$ query atoms. A conjunctive instance query $q$ w.r.t. a query signature $\mathcal{S}_q$ is a non-empty set of query atoms.*

*For a function $\mu$, we use $dom(\mu)$ to denote the domain of $\mu$. Let $\mathcal{O}$ be an ontology over $\mathcal{S}$ and $q = \{at_1, \ldots, at_n\}$ a query over $\mathcal{S}_q$ consisting of $n$ axiom templates. A mapping $\mu$ for $q$ over $\mathcal{O}$ is a total function $\mu \colon Vars(q) \to N_C^{\mathcal{O}} \cup N_R^{\mathcal{O}} \cup N_I^{\mathcal{O}}$ such that*

1. *$\mu(v) \in N_C^{\mathcal{O}}$ for each $v \in V_C \cap dom(\mu)$,*

2. *$\mu(v) \in N_R^{\mathcal{O}}$ for each $v \in V_R \cap dom(\mu)$,*

3. *$\mu(v) \in N_I^{\mathcal{O}}$ for each $v \in V_I \cap dom(\mu)$, and*

4. *$\mathcal{O} \cup \mu(q)$ is a $\mathcal{SROIQ}$ ontology.*

*We write $\mu(q)$ ($\mu(at)$) to denote the result of replacing each variable $v$ in $q$ (at) with $\mu(v)$. The set $\Gamma_q^{\mathcal{O}}$ of the compatible mappings for $q$ over $\mathcal{O}$ is defined as $\Gamma_q^{\mathcal{O}} := \{\mu \mid \mu \text{ is a mapping for } q \text{ over } \mathcal{O}\}$. A mapping $\mu$ is a solution mapping or a certain answer for $q$ over $\mathcal{O}$ if $\mathcal{O} \models \mu(q)$. We denote the set containing all solution mappings for $q$ over $\mathcal{O}$ with $\Omega_q^{\mathcal{O}}$. The result size or the number of answers of a query $q$ over $\mathcal{O}$ is given by the cardinality of the set $\Omega_q^{\mathcal{O}}$.*

Note that the last condition in the definition of mappings is required to ensure decidability of query entailment. For example, without the condition, a reasoner might have to test instantiated axiom templates where a role variable has been replaced by a non-simple role in a number restriction, which is not allowed in Description Logic axioms. Note also that we do not indicate which variables are to be selected since we do not consider the straightforward task of projection here.

Examples of queries according to the above definition are the following (where $?x$ is a concept variable, $?y$ a role variable, and $?z$ an individual variable):

$$C \sqsubseteq \exists ?y.?x$$
$$(\exists r.?x)(?z)$$

In the remainder, we use $\mathcal{S}$ for a signature $(N_C, N_R, N_I)$, $\mathcal{O}$ to denote a $\mathcal{SROIQ}$ ontology over $\mathcal{S}$, $A, B \in N_C$ for concept names from $\mathcal{O}$, $r, s \in N_R$ for role names from $\mathcal{O}$, $a, b \in N_I$ for individual names from $\mathcal{O}$, $?x, ?y$ for variables, $c_1, c_2$ for concept terms, $r_1, r_2$ for role terms, $t, t'$ for individual terms, $q = \{at_1, \ldots, at_n\}$ for a query with $n$ axiom templates over the query signature $\mathcal{S}_q = (N_C, N_R, N_I, V_C, V_R, V_I)$, $\Gamma_q^{\mathcal{O}}$ for the compatible mappings and $\Omega_q^{\mathcal{O}}$ for the solution mappings of $q$ over $\mathcal{O}$.

## 2.4 Model-building (Hyper)Tableau Calculi

In this section, we give a brief overview over the main reasoning techniques for OWL DL ontologies since our cost-based query planning relies on these techniques.

In order to check whether an ontology $\mathcal{O}$ entails an axiom $\alpha$, one typically checks whether $\mathcal{O} \cup \{\neg\alpha\}$ has a model. If that is not the case, then every model of $\mathcal{O}$ satisfies $\alpha$ and $\mathcal{O} \models \alpha$. For example, to check whether an individual $a_0$ is an instance of a concept $C$ w.r.t. an ontology $\mathcal{O}$, we check whether adding the concept assertion $\neg C(a_0)$ to $\mathcal{O}$ leads to an inconsistency. To check this, most OWL reasoners use a model construction calculus such as tableau or hypertableau. In the remainder, we focus on the hypertableau calculus (Motik, Shearer, & Horrocks, 2009), but a tableau calculus could equally be used and we state how our results can be transferred to tableau calculi.

The hypertableau calculus starts from the initial set of ABox assertions and, by applying derivation rules, it tries to construct (an abstraction of) a model of $\mathcal{O}$. Derivation rules usually add new concept or role assertion axioms, they may introduce new individuals, they can be nondeterministic, leading to the need to choose between several alternative assertion axioms to add or they can lead to a *clash* when a contradiction is detected. To show that an ontology $\mathcal{O}$ is (in)consistent, the hypertableau calculus constructs a *derivation*, i.e., a sequence of sets of assertions $\mathcal{A}_0, \ldots, \mathcal{A}_n$, such that $\mathcal{A}_0$ contains all ABox assertions in $\mathcal{O}$, $\mathcal{A}_{i+1}$ is the result of applying a derivation rule to $\mathcal{A}_i$ and $\mathcal{A}_n$ is the final set of assertions where no more rules are applicable. If a derivation exists such that $\mathcal{A}_n$ does not contain a clash, then $\mathcal{O}$ is consistent and $\mathcal{A}_n$ is called a *pre-model* of $\mathcal{O}$. Otherwise $\mathcal{O}$ is inconsistent. Each assertion in a set of assertions $\mathcal{A}_i$ is derived either deterministically or nondeterministically. An assertion is *derived deterministically* if it is derived by the application of a deterministic derivation rule from assertions that were all derived deterministically. Any other derived assertion is *derived nondeterministically*. It is easy to know whether an assertion was derived deterministically or not because of the dependency directed backtracking that most (hyper)tableau reasoners employ. In the pre-model, each individual $s_0$ is assigned a label $\mathcal{L}(s_0)$ representing the concepts it is (non)deterministically an instance of and each pair of individuals $\langle s_0, s_1 \rangle$ is assigned a label $\mathcal{L}(\langle s_0, s_1 \rangle)$ representing the roles through which individual $s_0$ is (non)deterministically related to individual $s_1$.

## 3. Motivation

A straightforward algorithm to compute the answers for a query $q$ is to test, for each mapping $\mu$, whether $\mathcal{O} \models \mu(q)$. Since only terms that are used in $\mathcal{O}$ can occur in the range of a mapping $\mu$ for $q$ over $\mathcal{O}$, there are finitely many mappings to test. In the worst case, however, the number of mappings that have to be tested is still exponential in the number of variables in the query. Such an algorithm is sound and complete if the reasoner used to decide entailment is sound and complete since we check all mappings for variables that can constitute actual solution mappings.

Optimizations cannot easily be integrated into the above sketched algorithm since it uses the reasoner to check for the entailment of the instantiated query as a whole and, hence, does not take advantage of relations or dependencies that may exist between the individual axiom templates in $q$. For a more optimized evaluation, one can evaluate the query axiom template by axiom template. Initially, the solution set contains only the identity mapping, which does not map any variable to a value. One then picks the first axiom template, extends the identity mapping to cover the variables of the chosen axiom template and then uses a reasoner to check which of the mappings instantiate the axiom template into an entailed axiom. One then picks the next axiom template and again extends the mappings from the previous round to cover all variables and checks which of those mappings lead to an entailed axiom. Thus, axiom templates which are very selective and are only satisfied by very few solutions reduce the number of intermediate solutions. Choosing a good execution order, therefore, can significantly affect the performance.

For example, let $q = \{A(?x), r(?x, ?y)\}$ with $?x, ?y \in V_I$. The query belongs to the class of conjunctive instance queries. We assume that the queried ontology contains 100 individuals, only 1 of which belongs to the concept $A$. This $A$ instance has 1 $r$-successor, while we have overall 200 pairs of individuals related with the role $r$. If we first evaluate $A(?x)$, we test 100 mappings (since $?x$ is an individual variable), of which only 1 mapping satisfies the axiom template. We then evaluate $r(?x, ?y)$ by extending the mapping with all 100 possible mappings for $?y$. Again only 1 mapping yields a solution. For the reverse axiom template order, the first axiom template requires the test of $100 \cdot 100$ mappings. Out of those, 200 remain to be checked for the second axiom template and we perform $10,200$ tests instead of just 200. Note also that the number of intermediate results when the query is evaluated in the order $A(?x)$, $r(?x, ?y)$ is smaller than when it is evaluated in the reverse order (2 versus 201).

In the context of databases or triple stores, cost-based ordering techniques for finding an optimal or near optimal join ordering have been widely applied (Steinbrunn, Moerkotte, & Kemper, 1997; Stocker, Seaborne, Bernstein, Kiefer, & Reynolds, 2008). These techniques involve the maintenance of a set of statistics about relations and indexes, e.g., number of pages in a relation, number of pages in an index, number of distinct values in a column, together with formulas for the estimation of the selectivity of predicates and the estimation of the CPU and I/O costs of query execution that depends amongst others, on the number of pages that have to be read from or written to secondary memory. The formulas for the estimation of selectivities of predicates (result output size of axiom templates) estimate the data distributions using histograms (Ioannidis & Christodoulakis, 1993), parametric or sampling methods or combinations of them. Ordering strategies as implemented in databases or triple stores are, however, not directly applicable in our setting. In the presence of expressive schema level axioms, we cannot rely on counting the number of occurrences of triples. We also cannot, in general, precompute all relevant inferences to base our statistics on materialized inferences. Furthermore, we should not only aim at decreasing the number of intermediate results, but also take into account the cost of checking or computing the solutions. This cost can be very significant with OWL reasoning and its precise estimation before query evaluation is difficult as this cost takes values from a wide range, e.g., due to nondeterminism and the high worst-case complexity of the standard reasoning tasks.[1]

For several kinds of axiom templates we can, however, directly retrieve the solutions from the reasoner instead of checking entailment. For example, for $C(?x)$, reasoners typically have a method to retrieve concept instances. Although this might internally trigger several tests, most methods of reasoners are highly optimized and avoid as many tests as possible. Furthermore, reasoners typically cache several results such as the computed concept hierarchy and retrieving sub-concepts can then be realized with a cache lookup. Thus, the actual execution cost might vary significantly. Notably, we do not have a straight correlation between the number of results for an axiom template and the actual cost of retrieving the solutions as is typically the case in triple stores or databases. This requires cost models that take into account the cost of the specific reasoning operations (depending on the state of the reasoner) as well as the number of results.

As motivated above, we distinguish between *simple* and *complex* axiom templates. Simple axiom templates are those that correspond to dedicated reasoning tasks. Let $c_1$ be a concept term, $C, C'$ (complex) concepts or concept variables, $r_1, r_2$ role terms or role inverses and $t, t'$ individual terms. The set of simple axiom templates contains templates of the form: $C \sqsubseteq C'$, $\exists r_1.\top \sqsubseteq c_1$ (domain restriction template), $\top \sqsubseteq \forall r_1.c_1$ (range restriction template), $r_1 \sqsubseteq r_2$, $C(t)$, $r_1(t, t')$, $t \approx t'$, $t \not\approx t'$. Complex axiom templates can, in contrast, not be evaluated by dedicated reasoning tasks and might require iterating over the compatible mappings and by checking entailment for each instantiated axiom template. An example of a complex axiom template is $(\exists r.?x)(?y)$.

## 4. Preprocessing for Extracting Information for Queries

In this section, we describe a way of preprocessing the queried ontology to extract information that is useful for ordering the axiom templates in a query. This preprocessing is useful for axiom templates of the form $c_1(t)$, $r_1(t, t')$, or $t \approx t'$, where $c_1$ is a concept term, $r_1$ is a role term and $t, t'$ are individual terms.

### 4.1 Extracting Individual Information from Reasoner Models

The first step in the ordering of query atoms is the extraction of statistics by exploiting information generated by reasoners. We use the labels of an initial pre-model to pro-

---

1. For example, the description logic $\mathcal{SROIQ}$, which underpins the OWL 2 DL standard, has a worst case complexity of 2-NExpTime (Kazakov, 2008) and typical implementations are not worst case optimal.

---

**Algorithm 1** initializeKnownAndPossibleConceptInstances($\mathcal{O}$)

---

**Input:** a consistent $\mathcal{SROIQ}$ ontology $\mathcal{O}$

1:   $\mathcal{A}_n := buildModelFor(\mathcal{O})$
2:  **for all** $a \in N_I^{\mathcal{O}}$ **do**
3:    **for all** $C \in \mathcal{L}_{\mathcal{A}_n}(a)$ **do**
4:      **if** $C$ was derived deterministically **then**
5:        $K[C] := K[C] \cup \{a\}$
6:      **else**
7:        $P[C] := P[C] \cup \{a\}$
8:      **end if**
9:    **end for**
10: **end for**

---

vide information about the concepts the individuals belong to or the roles with which one individual is connected to another one. We exploit this information similarly as was suggested for determining known or possible (non-)subsumers for concepts during classification (Glimm et al., 2012). In the hypertableau calculus, the following two properties hold for each ontology $\mathcal{O}$ and each constructed pre-model $\mathcal{A}_n$ for $\mathcal{O}$:

(P1) for each concept name $C$ (role name $r$), each individual $s_0$ (pair of individuals $\langle s_1, s_2 \rangle$) in $\mathcal{A}_n$, if $C \in \mathcal{L}_{\mathcal{A}_n}(s_0)$ $(r \in \mathcal{L}_{\mathcal{A}_n}(\langle s_1, s_2 \rangle))$ and the assertion $C(s_0)$ $(r(s_1, s_2))$ was derived deterministically, then it holds $\mathcal{O} \models C(s_0)$ $(\mathcal{O} \models r(s_1, s_2))$.

(P2) for an arbitrary individual $s_0$ in $\mathcal{A}_n$ (pair of individuals $\langle s_1, s_2 \rangle$ in $\mathcal{A}_n$) and an arbitrary concept name $C$ (simple role name $r$), if $C \notin \mathcal{L}_{\mathcal{A}_n}(s_0)$ $(r \notin \mathcal{L}_{\mathcal{A}_n}(\langle s_1, s_2 \rangle))$, then $\mathcal{O} \not\models C(s_0)$ $(\mathcal{O} \not\models r(s_1, s_2))$.

For simplicity, we assume here that equality ($\approx$) is axiomatized and $\approx$ is treated as a reflexive, symmetric, and transitive role. We use these properties to extract information from the pre-model of a satisfiable ontology $\mathcal{O}$.

**Definition 7** (**Known and Possible Instances**). *Let $\mathcal{A}_n$ be a pre-model for an ontology $\mathcal{O}$. An individual $a$ is a known (possible) instance of a concept name $C$ in $\mathcal{A}_n$, denoted $a \in K_{\mathcal{A}_n}[C]$ ($a \in P_{\mathcal{A}_n}[C]$), if $C \in \mathcal{L}_{\mathcal{A}_n}(a)$ and $C(a)$ is derived deterministically (nondeterministically) in $\mathcal{A}_n$. A pair of individuals $\langle a_1, a_2 \rangle$ is a known (possible) instance of a simple role name $r$ in $\mathcal{A}_n$, denoted $\langle a_1, a_2 \rangle \in K_{\mathcal{A}_n}(r)$, if $r \in \mathcal{L}_{\mathcal{A}_n}(\langle a_1, a_2 \rangle)$ and $r(a_1, a_2)$ is derived deterministically (nondeterministically) in $\mathcal{A}_n$. The individual $a_1$ is (possibly) equal to the individual $a_2$, written $a_1 \in K_{\approx}[a_2]$ and $a_2 \in K_{\approx}[a_1]$ ($a_1 \in P_{\approx}[a_2]$ and $a_2 \in P_{\approx}[a_1]$) if $a_1 \approx a_2$ has been deterministically (nondeterministically) derived in $\mathcal{O}$.*

In the remainder, we assume that the known and possible instances are defined w.r.t. some arbitrary pre-model $\mathcal{A}_n$ for $\mathcal{O}$ and we simply write $K[C]$, $K[r]$, $K_{\approx}[a]$, $P[C]$, $P[r]$, and $P_{\approx}[a]$. Intuitively, $K[C]$ contains individuals that can safely be considered instances of the concept name $C$. On the other hand, the possible instances require costly consistency checks in order to decide whether they are real instances of the concept, while individuals that neither belong to $K[C]$ nor $P[C]$ can safely be assumed to be non-instances of $C$.

Algorithm 1 outlines a procedure to initialize the relations for known and possible concept instances. The information we extract involves the maintenance of the sets of known and possible instances for all concepts of $\mathcal{O}$. One can define a similar algorithm for initializing the known and possible instances of simple roles and for (possibly) equal individuals. In our implementation, we use a more involved procedure to only store the direct types of each individual, where a concept name $C$ is a *direct type* of an individual $a$ in an ontology $\mathcal{O}$ if $\mathcal{O} \models C(a)$ and there is no concept name $D$ such that $\mathcal{O} \models D \sqsubseteq C$, $\mathcal{O} \models D(a)$ and $\mathcal{O} \not\models D \equiv C$.

Hypertableau and tableau reasoners typically do not deal with transitivity directly. In order to deal with non-simple roles, $\mathcal{O}$ is expanded with additional axioms that capture the semantics of the transitive relations before a pre-model is built. In particular, for each individual $a$ and non-simple role $r$, new concepts $C_a$ and $C_a^r$ are introduced and the axioms $C_a(a)$ and $C_a \sqsubseteq \forall r.C_a^r$ are added to $\mathcal{O}$. The consequent application of the transitivity encoding (Motik et al., 2009) produces axioms that propagate $C_a^r$ to each individual $b$ that is reachable from $a$ via an $r$-chain. The known and possible $r$-successors for $a$ can then be determined from the $C_a^r$ instances.

The technique presented in this paper can be used with any (hyper)tableau calculus for which properties (P1) and (P2) hold. All (hyper)tableau calculi used in practice that we are aware of satisfy property (P1). Pre-models produced by tableau algorithms as presented in the literature also satisfy property (P2); however, commonly used optimizations, such as lazy unfolding, can compromise property (P2), which we illustrate with the following example. Let us assume we have an ontology $\mathcal{O}$ containing the axioms

$$A \sqsubseteq \exists r.(C \sqcap D) \tag{1}$$

$$B \equiv \exists r.C \tag{2}$$

$$A(a) \tag{3}$$

It is obvious that in this ontology $A$ is a subconcept of $B$ (hence, $\mathcal{O} \models B(a)$) since every individual that is $r$-related to an individual that is an instance of the intersection of $C$ and $D$ is also $r$-related to an individual that is an instance of the concept $C$. However, even though the assertion $A(a)$ occurs in the ABox, the assertion $B(a)$ is not added in the pre-model when we use lazy unfolding. With lazy unfolding, instead of treating (2) as two disjunctions $\neg B \sqcup \exists r.C$ and $B \sqcup \forall r.(\neg C)$ as is typically done for general concept inclusion axioms, $B$ is only lazily unfolded into its definition $\exists r.C$ once $B$ occurs in the label of an individual. Thus, although $(\exists r.(C \sqcap D))(a)$ would be derived, this does not lead to the addition of $B(a)$.

Nevertheless, most (if not all) implemented calculi produce pre-models that satisfy at least the following weaker property:

(P3) for an arbitrary individual $s_0$ in $\mathcal{A}_n$ and an arbitrary concept name $C$ where $C$ is primitive in $\mathcal{O}$,[2] if $C \notin \mathcal{L}_{\mathcal{A}_n}(s_0)$, then $\mathcal{O} \not\models C(s_0)$.

Hence, properties (P2) and (P3) can be used to extract (non-)instance information from pre-models. For tableau calculi that only satisfy (P3), for each non-primitive concept name

---

2. A concept $C$ is considered primitive in $\mathcal{O}$ if $\mathcal{O}$ is unfoldable (Tsarkov et al., 2007) and it contains no axiom of the form $C \equiv E$ $C$ in $\mathcal{O}$ we need to add to $P[C]$ the individuals in $\mathcal{O}$ that do not include the concept $C$ in their label.

The proposed technique for determining known and possible instances of concept and role names can be used in the same way with both tableau and hypertableau reasoners. Since tableau algorithms often introduce more nondeterminism than hypertableau, one might, however, find less deterministic derivations, which results in less accurate statistics.

### 4.1.1 Individual Clustering

In this section, we describe the procedure for creating clusters of individuals within an ontology $\mathcal{O}$ using a constructed pre-model $\mathcal{A}_n$ of $\mathcal{O}$. Two types of clusters are created: *concept clusters* and *role clusters*. Concept clusters contain individuals having the same concepts in their label and role clusters contain individuals with the same concept and role labels. Role clusters are divided into three categories, those that are based on the first individual of role instances, those based on the second individual and those based on both individuals.

**Definition 8** (**Concept and Role Clusters**). *Let $\mathcal{O}$ be an ontology and $\mathcal{A}_n$ a pre-model for $\mathcal{O}$. We define the following two relations $P_1$ and $P_2$ that map an individual $a$ from $\mathcal{O}$ to the roles for which $a$ has at least one successor or predecessor, respectively:*

$$P_1(a) = \{r \mid r \in \mathcal{L}_{\mathcal{A}_n}(\langle a, b \rangle) \text{ for some } b \in N_I^{\mathcal{O}}\}$$
$$P_2(a) = \{r \mid r \in \mathcal{L}_{\mathcal{A}_n}(\langle b, a \rangle) \text{ for some } b \in N_I^{\mathcal{O}}\}$$

*Based on these relations, we build three different partitions over $N_I^{\mathcal{O}}$: concept clusters $CC$, role successor clusters $PC_1$, and role predecessor clusters $PC_2$ such that the clusters satisfy:*

*for each $C \in CC$.(for each $a_1, a_2 \in C.(\mathcal{L}_{\mathcal{A}_n}(a_1) = \mathcal{L}_{\mathcal{A}_n}(a_2)))$*
*for each $C \in PC_1$.(for each $a_1, a_2 \in C.(\mathcal{L}_{\mathcal{A}_n}(a_1) = \mathcal{L}_{\mathcal{A}_n}(a_2)$ and $P_1(a_1) = P_1(a_2)))$*
*for each $C \in PC_2$.(for each $a_1, a_2 \in C.(\mathcal{L}_{\mathcal{A}_n}(a_1) = \mathcal{L}_{\mathcal{A}_n}(a_2)$ and $P_2(a_1) = P_2(a_2)))$.*

*We further partition $N_I^{\mathcal{O}} \times N_I^{\mathcal{O}}$ into role clusters $PC_{12}$ such that the clusters satisfy:*

*for each $C \in PC_{12}$.(for each $\langle a_1, a_2 \rangle, \langle a_3, a_4 \rangle \in C.(\mathcal{L}_{\mathcal{A}_n}(a_1) = \mathcal{L}_{\mathcal{A}_n}(a_3), \mathcal{L}_{\mathcal{A}_n}(a_2) = \mathcal{L}_{\mathcal{A}_n}(a_4)$*
*and $\mathcal{L}_{\mathcal{A}_n}(\langle a_1, a_2 \rangle) = \mathcal{L}_{\mathcal{A}_n}(\langle a_3, a_4 \rangle)))$.*

We use these clusters in the next section to optimize the dynamic query ordering strategy.

## 5. Query Answering and Axiom Template Ordering

In this section, we describe two different algorithms (a static and a dynamic one) for ordering the axiom templates of a query based on some costs and then we deal with the formulation of these costs. We first introduce the abstract graph representation of a query $q$ by means of a labeled graph $G_q$ on which we define the computed statistical costs.

**Definition 9** (**Query Join Graph**). *A query join graph $G_q$ for a query $q$ is a tuple $(V, E, E_L)$, where*

- $V = q$ *is a set of vertices (one for each axiom template);*

- $E \subseteq V \times V$ *is a set of edges; such that* $\langle at_1, at_2 \rangle \in E$ *if* $Vars(at_1) \cap Vars(at_2) \neq \emptyset$ *and* $at_1 \neq at_2$;

- $E_L$ *is a function that assigns a set of variables to each* $\langle at_1, at_2 \rangle \in E$ *such that* $E_L(at_1, at_2) = Vars(at_1) \cap Vars(at_2)$.

In the remainder, we use $G_q$ for the query join graph of $q$.

Our goal is to find a query execution plan, which determines the evaluation order for axiom templates in $q$. Since the number of possible execution plans is of order $|q|!$, the ordering task quickly becomes impractical. In the following, we focus on greedy algorithms for determining an execution order, which prune the search space considerably. Roughly speaking, we proceed as follows: We define a cost function, which consists of two components (i) an estimate for the costs of the reasoning tasks needed for the evaluation of an axiom template and (ii) an estimate for the intermediate result size, i.e., the number of results that the evaluation of an axiom template will incur. Both components are combined to induce an order among axiom templates. In this paper, we simply build the sum of the two cost components, but different combinations such as a weighted sum of the two values could also be used. For the query plan construction we distinguish *static* from *dynamic planning*. For the former, we start constructing the plan by adding a minimal template according to the order. Variables from this template are then considered bound, which changes the cost function and might induce a different order among the remaining axiom templates. Considering the updated order, we again select the minimal axiom template that is not yet in the plan and update the costs. This process continues until the plan contains all templates. Once a complete plan has been determined the templates are evaluated. The dynamic case differs in that after selecting a template for the plan, we immediately determine the solutions for the chosen template, which are then used to update the cost function. While this yields accurate cost estimates, it can be very costly when all solutions are considered for updating the cost function. Sampling techniques can be used to only test a subset of the solutions, but we show in Section 7 that random sampling, i.e., randomly choosing a percentage of the individuals from the so far computed solutions, is not adequate. For this reason, we propose an alternative sampling approach that is based on the use of the previously described individual clusters. We first present an example to make the difference between static and dynamic planning clearer and justify why dynamic ordering can be beneficial in our setting.

**Example 1.** *Let* $\mathcal{O}$ *be an ontology and* $q = \{C(?x), r(?x, ?y), D(?y)\}$ *a conjunctive instance query over* $\mathcal{O}$. *Suppose that for the known and possible instances of the query concepts and roles we have*

$$K[C] = \{a\} \qquad K[r] = \emptyset \qquad K[D] = \{b\}$$
$$P[C] = \{c, e\} \qquad P[r] = \{\langle c, d \rangle, \langle e, f \rangle\} \qquad P[D] = \{f, g, h\}$$

*And let us assume that the possible instances of* $C$, $D$ *and* $r$ *are, in fact, real instances (note that we do not have this information from the beginning). Please have in mind that the possible instances of concepts or roles are more costly to evaluate than the known instances* *since they require expensive consistency checks in order to decide whether they are real instances.*

*According to static planning, an ordering for query atoms is first determined. In particular, the atom $r(?x, ?y)$ is chosen first since it has the least number of known and possible instances (0 known and 2 possible versus 1 known and 2 possible for $C(?x)$ and 1 known and 3 possible for $D(?y)$). Then the atom $C(?x)$ is chosen since it has less known and possible instances than $D(?y)$, i.e., 1 known and 2 possible versus 1 known and 3 possible for $D(?y)$. Hence the chosen execution plan in static planning is $P = (r(?x, ?y), C(?x), D(?y))$. Afterwards, the query is evaluated according to the chosen execution plan, i.e., the atom $r(?x, ?y)$ is evaluated first, which gives the solution mappings $\Omega_1 = \{\{?x \mapsto c, ?y \mapsto d\}, \{?x \mapsto e, ?y \mapsto f\}\}$. This requires 2 consistency checks for the 2 possible instances of $r$. Afterwards, we check which of the $?x$ mappings, $c$ and $e$, are known or possible instances of $C$. Since both $c$ and $e$ are possible instances, we check whether they are real instances of $C$ (this requires 2 consistency checks). Hence, the solution mappings are $\Omega_2 = \Omega_1 = \{\{?x \mapsto c, ?y \mapsto d\}, \{?x \mapsto e, ?y \mapsto f\}\}$. In the end, we check which of the $?y$ mappings, $d$ and $f$, are known or possible instances of $D$. For the only possible instance, $f$, we find after one consistency check that $f$ is indeed an instance of $D$. Hence, the solution mappings for $q$ are $\Omega_q^{\mathcal{O}} = \{\{?x \mapsto e, ?y \mapsto f\}\}$ and finding the solution required 5 consistency checks.*

*According to dynamic planning, an ordering is determined while we evaluate the query. For the same reasons as before, the atom $r(?x, ?y)$ is chosen to be evaluated first and the solution mappings are, as before, $\Omega_1 = \{\{?x \mapsto c, ?y \mapsto d\}, \{?x \mapsto e, ?y \mapsto f\}\}$ (this requires 2 consistency checks). We afterwards check which of the $?y$ mappings, $d$ and $f$, are known or possible instances of $D$. Note that this only requires a look-up since if we find $d$ or $f$ to be among the possible instances, we do not check whether the individual is indeed an instance or not. Here only $f$ is a possible instance. We also check which of the $?x$ mappings, $c$ and $e$, are known or possible instances of $C$. Here, both $c$ and $e$ are possible instances, i.e., we have 2 relevant possible instances for $C(?x)$ and 1 for $D(?y)$. Hence, the atom $D(?y)$ is chosen to be evaluated next, resulting in the solution sequence $\Omega_2 = \{\{?x \mapsto e, ?y \mapsto f\}\}$ for the (partial) execution plan $(r(?x, ?y), D(?y))$, requiring 1 consistency check. In the end, we check whether the $?x$ mapping, $e$, is a known or possible instance of $C$. Since $e$ is a possible instance, we check whether it is a real instance (this requires 1 consistency check). Hence, the solution mappings for $q$ are $\Omega_q^{\mathcal{O}} = \{\{?x \mapsto e, ?y \mapsto f\}\}$, which have been found by performing 4 consistency checks, one less than in the static case.*

*Note that in dynamic ordering we perform less checks than in static ordering, since in this case we can exploit the results of joins of query atoms and more information regarding the possible instances of atoms (i.e., which of them are real instances), which is determined as a result of evaluating the atoms while ordering them.*

We now make the process of query plan construction more precise, but we leave the exact details of defining the cost function and the ordering it induces to later.

**Definition 10 (Static and Dynamic Ordering).** *A* static (dynamic) cost function w.r.t. $q$ over $\mathcal{O}$ *is a function $s: q \times 2^{Vars(q)} \to \mathbb{R} \times \mathbb{R}$ ($d: q \times 2^{\Gamma_q^{\mathcal{O}}} \to \mathbb{R} \times \mathbb{R}$), where with $\Gamma_q^{\mathcal{O}}$ we denote the set of compatible mappings for $q$ over $\mathcal{O}$. The two costs $\langle \mathsf{Ec}_{at}^s, \mathsf{Rs}_{at}^s \rangle$ ($\langle \mathsf{Ec}_{at}^d, \mathsf{Rs}_{at}^d \rangle$) for an axiom template $at \in q$ are combined to yield a static ordering $\preceq_s$ (dynamic ordering* $\preceq_d$), which is a total order over the axiom templates of $q$ such that, for $at, at' \in q$, we say that $at \preceq_s at'$ ($at \preceq_d at'$) iff $\mathsf{Ec}^s_{at} + \mathsf{Rs}^s_{at} \leq \mathsf{Ec}^s_{at'} + \mathsf{Rs}^s_{at'}$ ($\mathsf{Ec}^d_{at} + \mathsf{Rs}^d_{at} \leq \mathsf{Ec}^d_{at'} + \mathsf{Rs}^d_{at'}$).

An execution plan *for $q$ is a duplicate-free sequence of axiom templates from $q$. The* initial execution plan *is the empty sequence and a* complete execution plan *is a sequence containing all templates of $q$. Let $P_i = (at_1, \ldots, at_i)$ with $i < |q|$ be an execution plan for $q$ with query join graph $G_q = (V, E, E_L)$. The set of* bound variables *of $at_i$ within $P_i$ is $V_b(at_i) = Vars(at_i) \cap Vars(\{at_1, \ldots, at_{i-1}\})$. Let $C_q$ be the set of complex axiom templates in $q$. We next define which axiom templates can be used to extend an incomplete execution plan. Let $at$ be an axiom template in $P_i$, the set $suc_i(at)$ contains the axiom templates that are connected to $at$ and not yet in $P_i$, i.e., $suc_i(at) = \{at' \in q \mid \langle at, at' \rangle \in E, at' \notin \{at_1, \ldots at_i\}\}$. Based on this, we define the set of* connected successor axiom templates *for $P_i$ as $S_i = \{at \mid at' \in \{at_1, \ldots, at_i\} \text{ and } at \in suc_i(at')\}$. We further allow for including axiom templates that are only connected to a complex axiom template from $S_i$ and define the* potential next templates *$q_i$ for $P_i$ w.r.t. $G_q$ as $q_i = q$ if $P_i$ is the initial execution plan and otherwise*

$$q_i = S_i \cup \bigcup_{at \,\in\, C_q \,\cap\, S_i} suc_i(at).$$

*Given $P_i$, the static (dynamic) ordering induces an execution plan $P_{i+1} = (at_1, \ldots, at_i, at_{i+1})$ with $at_{i+1} \in q_i$ and $at_{i+1} \preceq_s at$ ($at_{i+1} \preceq_d at$) for each $at \in q_i$ such that $at \neq at_{i+1}$.*

Note that according to the above definition, for $P_i$ an execution plan, it can be the case that $q_i$ contains templates that are assigned the same minimal cost by the cost function. In such case, one can choose any of these atoms to add to $P_i$. Moreover, according to the above definition for the case of queries containing only simple axiom templates we have that, for $i > 0$, the set of potential next templates only contains templates that are connected to a template that is already in the plan since unconnected templates cause an unnecessary blowup of the number of intermediate results. For queries with complex templates the set of potential next axiom templates can additionally contain templates that do not share common variables with any template that is already in the plan. This different handling of queries with complex templates is reasonable since, before evaluating a complex axiom template that requires many consistency checks, we want to reduce the number of candidate bindings, by first evaluating other simple (cheaper) templates that bind variables which appear in the complex one.

**Example 2.** *Let $\mathcal{O}$ be an ontology and $q = \{?x \sqsubseteq A, ?y \sqsubseteq r, B \sqsubseteq \exists ?y.?x\}$ a query over $\mathcal{O}$. Assuming that systems usually precompute the concept and role hierarchies before they accept queries, the evaluation of the first two templates, i.e., $?x \sqsubseteq A$ and $?y \sqsubseteq r$, require cheap cache lookups, whereas the axiom template $B \sqsubseteq \exists ?y.?x$, requires costly consistency checks. Hence, it is reasonable to first evaluate the first two (cheap) templates to reduce the mappings for $?x$ and $?y$ and then evaluate the third (expensive) template, by checking which of the reduced mappings yield an entailed axiom.*

An example that shows the actual gain we get from handling the ordering of complex axiom templates in this way is presented in Section 7.

Let $n = |q|$ and $P_n = (at_1, \ldots, at_n)$ be a complete execution plan for $q$ over $\mathcal{O}$ determined by static ordering. The procedure to find the solution mappings $\Omega^{\mathcal{O}}_q$ for $P_n$ is recursively defined as follows: Initially, our solution set contains only the identity mapping $\Omega_0 = \{\mu_0\}$, which does not map any variable to any value. Assuming that we have evaluated the sequence $P_i = (at_1, \ldots, at_i)$, $i < n$ and we have found the set of solution mappings $\Omega_i$, in order to find the solution mappings $\Omega_{i+1}$ of $P_{i+1}$, we use specific reasoning tasks to extend the mappings in $\Omega_i$ to cover the new variables of $at_{i+1}$ if $at_{i+1}$ is a simple axiom template or the entailment check service of reasoners if $at_{i+1}$ does not contain new variables or if $at_{i+1}$ is a complex axiom template. In dynamic planning the difference is that the execution plan construction is interleaved with query evaluation. In particular, let $n = |q|$ and $P_i = (at_1 \ldots at_i)$ with $i < n$ be a (partial) execution plan for $q$ determined by dynamic ordering and let $\Omega_i$ be the solution mappings of $P_i$. In order to find $P_{i+1}$ we extend $P_i$ with a new template, $at_{i+1}$, from $q$, i.e., $P_{i+1} = (at_1, \ldots at_{i+1})$, which, according to the dynamic cost function, has the minimal cost among the remaining templates $q \setminus \{at_1, \ldots at_i\}$. The dynamic cost function assigns costs to templates at iteration $i + 1$ taking into account the solution mappings $\Omega_i$. We afterwards evaluate the atom $at_{i+1}$, i.e., we find the solution mappings $\Omega_{i+1}$ of $P_{i+1}$ by extending the solution mappings $\Omega_i$ of $P_i$ in the same way as in the static case. In Section 6.3 in Algorithm 3, we show the complete procedure we follow to answer a query.

We now define the cost functions $s$ and $d$ more precisely, which estimate the cost of the required reasoner operations (first component) and the estimated result output size (second component) of evaluating an axiom template. The intuition behind the estimated value of the reasoner operation costs is that the evaluation of possible instances is much more costly than the evaluation of known instances since possible instances require expensive consistency checks whereas known instances require cheap cache lookups. The estimated result size takes into account the number of known and possible instances and the probability that possible instances are actual instances.

The time needed for an entailment check can change considerably between ontologies and even within an ontology (depending on the involved concepts, roles and individuals). In order to more accurately determine the entailment cost we use different entailment cost values depending on whether the template under consideration is a template of the form i) $c_1(t)$, ii) $r_1(t, t')$, iii) $t \approx t'$, where $c_1$ is a concept term, $r_1$ is a role term and $t, t'$ are individual terms, iv) one of the rest simple axiom templates (that require consistency checks to be evaluated) or a complex axiom template. In the following we write $C_L$ to denote the cost of a cache lookup in the internal structures of the reasoner, $C_E$ as a placeholder for the relevant entailment cost value and $P_{IS}$ for the possible instance success, i.e, the estimated percentage of possible instances that are actual instances. The costs $C_L$ and $C_E$ are determined by recording the average time of previously performed lookups and entailment checks for the queried ontology, e.g., during the initial consistency check, classification, or for previous queries. The possible instance success, $P_{IS}$, was determined by testing several ontologies and checking how many of the initial possible instances were real ones, which was around 50% in nearly all ontologies.

Apart from the relations for the known and possible instances from Section 4.1, we use the following auxiliary relations:

**Definition 11 (Successor and Predecessor Relations).** *Let $r$ be a role and $a$ an individual. We define* sucK$[r]$ *and* preK$[r]$ *as the set of individuals with known $r$-successors and* *r-predecessors, respectively:*

$$\mathsf{sucK}[r] := \{a \mid \exists b. \langle a, b \rangle \in K[r]\} \qquad and \qquad \mathsf{preK}[r] := \{a \mid \exists b. \langle b, a \rangle \in K[r]\}.$$

*Similarly, we define* $\mathsf{sucK}[r, a]$ *and* $\mathsf{preK}[r, a]$ *as the known r-successors of a and the known r-predecessors of a, respectively:*

$$\mathsf{sucK}[r, a] := \{b \mid \langle a, b \rangle \in K[r]\} \qquad and \qquad \mathsf{preK}[r, a] := \{b \mid \langle b, a \rangle \in K[r]\}.$$

*We analogously define the functions* $\mathsf{sucP}[r]$, $\mathsf{preP}[r]$, $\mathsf{sucP}[r, a]$, *and* $\mathsf{preP}[r, a]$ *by replacing* $K[r]$ *with* $P[r]$.

Next, we define the cost functions for the case of conjunctive instance queries, i.e., queries containing only query atoms. In Section 5.2 we extend the cost functions to deal with general queries.

### 5.1 The Static and Dynamic Cost Functions for Conjunctive Instance Queries

The static cost function $s$ takes two components as input: a query atom and a set containing the variables of the query atom that are considered bound. The function returns a pair of real numbers for the reasoning cost and the result size for the query atom.

Initially, all variables are unbound and we use the number of known and possible instances or successors/predecessors to estimate the number of required lookups and consistency checks for evaluating the query atom and for the resulting number of mappings. For an input of the form $\langle C(?x), \emptyset \rangle$ or $\langle r(?x, ?y), \emptyset \rangle$ the resulting pair of real numbers for the computational cost and the estimated result size is computed as

$$\langle |K[at]| \cdot d \cdot C_L + |P[at]| \cdot d \cdot C_E, |K[at]| + P_{IS} \cdot |P[at]| \rangle,$$

where $at$ denotes the predicate of the query atom ($C$ or $r$). For $at$ a concept (role) atom, the factor $d$ represents the depth of the concept (role) in the concept (role) hierarchy. We use this factor since we only store the direct types of each individual (roles of which individuals are instances) and, in order to find the instances of a concept (role), we may need to check all its subconcepts (subroles) for known or possible instances. If the query atom is a role atom with a constant in the first place, i.e., the input to the cost function is of the form $\langle r(a, ?x), \emptyset \rangle$, we use the relations for known and possible successors to estimate the computational cost and result size:

$$\langle |\mathsf{sucK}[r, a]| \cdot d \cdot C_L + |\mathsf{sucP}[r, a]| \cdot d \cdot C_E, |\mathsf{sucK}[r, a]| + P_{IS} \cdot |\mathsf{sucP}[r, a]| \rangle.$$

Analogously, we use $\mathsf{preK}$ and $\mathsf{preP}$ instead of $\mathsf{sucK}$ and $\mathsf{sucP}$ for an input of the form $\langle r(?x, a), \emptyset \rangle$. Finally, if the atom contains only constants, i.e., the input to the cost function is of the form $\langle C(a), \emptyset \rangle, \langle r(a, b), \emptyset \rangle$, the function returns $\langle d \cdot C_L, 1 \rangle$ if the individual is a known instance of the concept or role, $\langle d \cdot C_E, P_{IS} \rangle$ if the individual is a possible instance and $\langle d \cdot C_L, 0 \rangle$ otherwise, i.e., if the individual is a known non-instance.

For equality atoms of the form $?x \approx ?y$, $a \approx ?x$, $?x \approx a$ or $a \approx b$, we again exploit information from the initial pre-model as described in Section 4.1. Based on the cardinality of $K_{\approx}[a]$ and $P_{\approx}[a]$, we can define cost functions for the different cases of query atoms and bound variables. For inputs of the form $\langle ?x \approx a, \emptyset \rangle$ and $\langle a \approx ?x, \emptyset \rangle$, the cost function is defined as:

$$\langle |K_{\approx}[a]| \cdot C_L + |P_{\approx}[a]| \cdot C_E, |K_{\approx}[a]| + P_{IS} \cdot |P_{\approx}[a]| \rangle.$$

For inputs of the form $\langle ?x \approx ?y, \emptyset \rangle$, the cost function is computed as:

$$\left\langle \sum_{a \in N_I^{\mathcal{O}}} (|K_{\approx}[a]| \cdot C_L + |P_{\approx}[a]| \cdot C_E)/2, \sum_{a \in N_I^{\mathcal{O}}} (|K_{\approx}[a]| + P_{IS} \cdot |P_{\approx}[a]|)/2 \right\rangle.$$

For inputs of the form $\langle a \approx b, \emptyset \rangle$, the function returns $\langle C_L, 1 \rangle$ if $b \in K_{\approx}[a]$, $\langle C_E, P_{IS} \rangle$ if $b \in P_{\approx}[a]$, and $\langle C_L, 0 \rangle$ otherwise (i.e., $b$ is not equivalent to $a$).

After determining the cost of an initial query atom, at least one variable of a consequently considered atom is bound, since during the query plan construction we move over atoms sharing a common variable and we assume that the query is connected. We now define the cost functions for atoms with at least one variable bound. We make the assumption that atoms with unbound variables are more costly to evaluate than atoms with all their variables bound. For a query atom $r(?x, ?y)$ with only $?x$ bound, i.e., function inputs of the form $\langle r(?x, ?y), \{?x\} \rangle$, we use the average number of known and possible successors of the role to estimate the computational cost and result size:

$$\left\langle \frac{|K[r]|}{|\mathsf{sucK}[r]|} \cdot d \cdot C_L + \frac{|P[r]|}{|\mathsf{sucP}[r]|} \cdot d \cdot C_E, \frac{|K[r]|}{|\mathsf{sucK}[r]|} + \frac{|P[r]|}{|\mathsf{sucP}[r]|} \cdot P_{IS} \right\rangle.$$

In case only $?y$ in $r(?x, ?y)$ is bound, we use the predecessor functions $\mathsf{preK}$ and $\mathsf{preP}$ instead of $\mathsf{sucK}$ and $\mathsf{sucP}$. Note that we now work with an estimated average number of successors (predecessors) for *one individual*.

For atoms with all their variables bound, we use formulas that are comparable to the ones above for an initial plan, but normalized to estimate the values for one individual. For an input query atom of the form $C(?x)$ with $?x$ a bound variable we use

$$\left\langle \frac{|K[C]| \cdot d \cdot C_L + |P[C]| \cdot d \cdot C_E}{|N_I^{\mathcal{O}}|}, \frac{|K[C]| + P_{IS} \cdot |P[C]|}{|N_I^{\mathcal{O}}|} \right\rangle.$$

Such a simple normalization is not always accurate, but leads to good results in most cases as we show in Section 7. Similarly, we normalize the formulas for role atoms of the form $r(?x, ?y)$ such that $\{?x, ?y\}$ is the set of bound variables of the atom. The two cost components for these atoms are computed as

$$\left\langle \frac{|K[r]| \cdot d \cdot C_L + |P[r]| \cdot d \cdot C_E}{|N_I^{\mathcal{O}}| \cdot |N_I^{\mathcal{O}}|}, \frac{|K[r]| + P_{IS} \cdot |P[r]|}{|N_I^{\mathcal{O}}| \cdot |N_I^{\mathcal{O}}|} \right\rangle.$$

For role atoms with a constant and a bound variable, i.e., atoms of the form $r(a, ?x)$ ($r(?x, a)$) with $?x$ a bound variable, we use $\mathsf{sucK}[r, a]$ and $\mathsf{sucP}[r, a]$ ($\mathsf{preK}[r, a]$ and $\mathsf{preP}[r, a]$) instead of $K[r]$ and $P[r]$ in the above formulas and we normalize by $|N_I^{\mathcal{O}}|$.

Similarly, we normalize the cost functions for inputs with equality atoms and bound variables, depending on whether the atoms contain one or two bound variables. For inputs of the form $\langle ?x \approx a, \{?x\} \rangle$, $\langle a \approx ?x, \{?x\} \rangle$, we divide the cost function components for inputs of the form $\langle ?x \approx a, \emptyset \rangle$ and $\langle a \approx ?x, \emptyset \rangle$ by $|N_I^{\mathcal{O}}|$. For an input of the form $\langle ?x \approx y, \{?x, ?y\} \rangle$, we divide the cost function components for input of the form $\langle ?x \approx ?y, \emptyset \rangle$ by $|N_I^{\mathcal{O}}| \cdot |N_I^{\mathcal{O}}|$. For inputs of the form $\langle ?x \approx ?y, \{?x\} \rangle$, and $\langle ?x \approx ?y, \{?y\} \rangle$, we divide the cost function components for input of the form $\langle ?x \approx ?y, \emptyset \rangle$ by $|N_I^{\mathcal{O}}|$.

| | already executed | current atom $at$ | $K[at]$ | $P[at]$ | real from $P[at]$ |
|---|---|---|---|---|---|
| 1 | | $C(?x)$ | 200 | 350 | 200 |
| 2 | | $r(?x,?y)$ | 200 | 200 | 50 |
| 3 | | $D(?y)$ | 700 | 600 | 400 |
| 4 | $r(?x,?y)$ | $C(?x)$ | 100 | 150 | 100 |
| 5 | $r(?x,?y)$ | $D(?y)$ | 50 | 50 | 40 |
| 6 | $r(?x,?y)$, $D(?y)$ | $C(?x)$ | 45 | 35 | 25 |
| 7 | $r(?x,?y)$, $C(?x)$ | $D(?y)$ | 45 | 40 | 25 |

Table 1: Query Ordering Example

The dynamic cost function $d$ is based on the static function $s$, but only uses the first equations, where the atom contains only unbound variables or constants. The function takes a pair $\langle at, \Omega \rangle$ as input, where $at$ is a query atom and $\Omega$ is the set of solution mappings for the atoms that have already been evaluated, and returns a pair of real numbers using matrix addition as follows:

$$d(at, \Omega) = \sum_{\mu \in \Omega} s(\mu(at), \emptyset)$$

When sampling techniques are used, we compute the costs for each of the potential next atoms for an execution plan by only considering one individual of each relevant cluster. Which cluster is relevant depends on the query atom for which we compute the cost function and the previously computed bindings. For instance, if we compute the cost of a role atom $r(?x,?y)$ and we have already determined bindings for $?x$, we use the role successor cluster $PC_1$. Among the $?x$ bindings, we then just check the cost for one binding per cluster and assign the same cost to all other $?x$ bindings of the same cluster.

**Example 3.** *Let us assume that we have a conjunctive instance query $q$ and that we have to find the cost (using the dynamic function) of the atom $C(?x)$ within an execution plan for $q$. We further assume that from the evaluation of previous query atoms in the plan we have already determined a set of intermediate solutions $\Omega$ with the mappings $a, b,$ or $c$ for $?x$. Let us assume that $a, b,$ and $c$ belong to the same concept cluster. According to dynamic ordering we need to find the cost of each instantiated atom using the static cost function, i.e., $d(C(?x), \Omega) = s(C(a), \emptyset) + s(C(b), \emptyset) + s(C(c), \emptyset)$. If we additionally use cluster based sampling, we find the cost for only one individual of each cluster, let us say $a$, and then assign the same cost to all other individuals from the cluster which are mappings for $?x$ in $\Omega$. Hence, the cost of the atom $C(?x)$ when sampling is used, is computed as $d(C(?x), \Omega) = 3 \cdot s(C(a), \emptyset)$ avoiding the computation of $s(C(b), \emptyset)$ and $s(C(c), \emptyset)$.*

An example that is similar to Example 1 (but with a greater number of instances) and shows how ordering is achieved by the use of the defined static and dynamic functions is shown below. We assume that $q$ is a query consisting of the three query atoms: $C(?x)$, $r(?x, ?y)$, $D(?y)$. Table 1 gives information about the known and possible instances of these atoms within a sequence. The second column shows already executed sequences $P_{i-1} = (at_1, \ldots, at_{i-1})$ for the atoms of $q$. Column 3 gives the current atom $at_i$ and column 4 (5) gives the number of mappings to known (possible) instances of $at$ that satisfy at the same time the atoms $(at_1, \ldots, at_{i-1})$ from column 2. Column 6 gives the number of real instances from the possible instances for the current atom. For example, row 4 says that we have evaluated the atom $r(?x, ?y)$ and, in order to evaluate $C(?x)$, we only consider those 100 known and 150 possible instances of $C$ that are also mappings for $?x$. We further assume that we have 10,000 individuals in our ontology $\mathcal{O}$. We now explain, using the example, how the above described formulas work. We assume that $C_L \leq C_E$, which is always the case since a cache lookup is less expensive than a consistency check and that the $C_E$ values are the same for all query concepts and roles. For ease of presentation, we further do not consider the factor for the depth of the concept (role) within the concept (role) hierarchy. In both techniques (static and dynamic) the atom $r(?x, ?y)$ is chosen first since it has the least number of possible instances (200) while it has the same (or smaller) number of known instances (200) as the other atoms ($\mu_0$ is the initial solution mapping that does not map any variable):

$$s(r(?x, ?y), \emptyset) = d(r(?x, ?y), \{\mu_0\}) = \langle 200 \cdot C_L + 200 \cdot C_E, 200 + P_{IS} \cdot 200 \rangle,$$
$$s(C(?x), \emptyset) = d(C(?x), \{\mu_0\}) = \langle 200 \cdot C_L + 350 \cdot C_E, 200 + P_{IS} \cdot 350 \rangle,$$
$$s(D(?y), \emptyset) = d(D(?y), \{\mu_0\}) = \langle 700 \cdot C_L + 600 \cdot C_E, 700 + P_{IS} \cdot 600 \rangle.$$

In the case of static ordering, the atom $C(?x)$ is chosen after $r(?x, ?y)$ since $C$ has less possible (and known) instances than $D$ (350 versus 600):

$$s(C(?x), \{?x\}) = \left\langle \frac{200}{10,000} \cdot C_L + \frac{350}{10,000} \cdot C_E, \frac{200 + 350 \cdot P_{IS}}{10,000} \right\rangle,$$
$$s(D(?y), \{?y\}) = \left\langle \frac{700}{10,000} \cdot C_L + \frac{600}{10,000} \cdot C_E, \frac{700 + 600 \cdot P_{IS}}{10,000} \right\rangle.$$

Hence, the order of evaluation in this case is $P = (r(?x, ?y), C(?x), D(?y))$ leading to 200 (row 2) + 150 (row 4) + 40 (row 7) entailment checks. In the dynamic case, after the evaluation of $r(?x, ?y)$, which gives a set of solutions $\Omega_1$, the atom $D(?y)$ has fewer known and possible instances (50 known and 50 possible) than the atom $C(?x)$ (100 known and 150 possible) and, hence, a lower cost:

$$d(D(?y), \Omega_1) = \langle 50 \cdot C_L + 150 \cdot C_L + 50 \cdot C_E, 50 + 0 + 50 \cdot P_{IS} \rangle,$$
$$d(C(?x), \Omega_1) = \langle 100 \cdot C_L + 0 \cdot C_L + 150 \cdot C_E, 100 + 0 + 150 \cdot P_{IS} \rangle.$$

Note that applying a solution $\mu \in \Omega_1$ to $D(?y)$ ($C(?x)$) results in a query atom with a constant in place of $?y$ ($?x$). For $D(?y)$, it is the case that out of the 250 $r$-instances, 200 can be handled with a look-up (50 turn out to be known instances and 150 turn out not to be instances of $D$), while 50 require an entailment check. Similarly, when considering $C(?x)$, we need 100 lookups and 150 entailment checks. Note that we assume the worst case in this example, i.e., that all values that $?x$ and $?y$ take are different. Therefore, the atom $D(?y)$ is chosen next, leading to the execution of the query atoms in the order $P = (r(?x, ?y), D(?y), C(?x))$ and the execution of 200 (row 2) + 50 (row 5) + 35 (row 6) entailment checks.

## 5.2 Cost Functions for General Queries

We now explain how we order the remaining simple and complex axiom templates. We again use statistics from the reasoner, whenever these are available. In case the reasoner cannot give estimates, one can still work with statistics computed from explicitly stated information or use upper bounds to estimate the reasoner costs and the result size of axiom templates.

We first consider a general concept assertion axiom template. Let $K_C[a]$ be the concepts of which $a$ is a known instance, $P_C[a]$ the concepts of which $a$ is a possible instance. These sets are computed from the sets of known and possible instances of concepts. For an input of the form $\langle ?x(a), \emptyset \rangle$ the cost function is defined as

$$\langle |K_C[a]| \cdot d \cdot C_L + |P_C[a]| \cdot d \cdot C_E, |K_C[a]| + P_{IS} \cdot |P_C[a]| \rangle.$$

For an input of the form $\langle ?x(?y), \emptyset \rangle$, the cost function is defined as

$$\left\langle \sum_{C \in N_C^O} (|K[C]| \cdot d \cdot C_L + |P[C]| \cdot d \cdot C_E), \sum_{C \in N_C^O} (|K[C]| + P_{IS} \cdot |P[C]|) \right\rangle.$$

For inputs of the form $\langle ?x(a), \{?x\} \rangle$ and $\langle ?x(?y), \{?x, ?y\} \rangle$, we normalize the above functions by $|N_C^O|$ and $|N_I^O| \| N_C^O|$ respectively. For inputs of the form $\langle ?x(?y), \{?x\} \rangle$ and $\langle ?x(?y), \{?y\} \rangle$ we normalize the function for inputs of the form $\langle ?x(?y), \emptyset \rangle$ by $|N_C^O|$ and $|N_I^O|$ respectively.

For general role assertion axiom templates, there are several cases of cost functions depending on the bound variables. We next define the cost functions for some cases. The cost functions for the other cases can similarly be defined. For an input of the form $\langle ?z(?x, ?y), \emptyset \rangle$, the cost function is defined as :

$$\left\langle \sum_{r \in N_R^O} (|K[r]| \cdot d \cdot C_L + |P[r]| \cdot d \cdot C_E), \sum_{r \in N_R^O} (|K[r]| + P_{IS} \cdot |P[r]|) \right\rangle.$$

For inputs of the form $\langle ?z(a, ?y), \emptyset \rangle$, the cost function is defined as:

$$\left\langle \sum_{r \in N_R^O} (\mathsf{sucK}[r, a] \cdot d \cdot C_L + \mathsf{sucP}[r, a] \cdot d \cdot C_E), \sum_{r \in N_R^O} (\mathsf{sucK}[r, a] + P_{IS} \cdot \mathsf{sucP}[r, a]) \right\rangle.$$

For an input of the form $\langle ?z(?x, ?y), \{?z\} \rangle$, the cost function is defined as:

$$\left\langle \sum_{r \in N_R^O} \frac{|K[r]| \cdot d \cdot C_L + |P[r]| \cdot d \cdot C_E}{|N_R^O|}, \sum_{r \in N_R^O} \frac{|K[r]| + P_{IS} \cdot |P[r]|}{|N_R^O|} \right\rangle.$$

Last, for inputs of the form $\langle ?z(?x, ?y), \{?x\} \rangle$, the two cost components are computed as:

$$\left\langle \sum_{r \in N_R^O} \left( \frac{|K[r]|}{|\mathsf{sucK}[r]|} \cdot d \cdot C_L + \frac{|P[r]|}{|\mathsf{sucP}[r]|} \cdot d \cdot C_E \right), \sum_{r \in N_R^O} \left( \frac{|K[r]|}{|\mathsf{sucK}[r]|} + \frac{|P[r]|}{|\mathsf{sucP}[r]|} \cdot P_{IS} \right) \right\rangle.$$

For concept (role) inclusion axiom templates of the form $c_1 \sqsubseteq c_2$ ($r_1 \sqsubseteq r_2$), where $c_1, c_2$ concept terms ($r_1, r_2$ role terms), that contain only concept (role) names and variables we need lookups in the computed concept (role) hierarchy in order to compute the answers (assuming that the concept (role) hierarchy is precomputed).

One can define similar cost functions for other types of axiom templates by either using the available statistics or by relying on told information from the ontology. For this paper, however, we just define a cost function based on the assumption that we iterate over all possible values of the respective variables and do one consistency check for each value. Hence, we define the following general cost function for these cases:

$$\langle |N| \cdot C_E, |N| \rangle,$$

where $N \in \{N_C^{\mathcal{O}}, N_R^{\mathcal{O}}, N_I^{\mathcal{O}}\}$ as appropriate for the variable that is tested. As discussed in Section 5.1, the dynamic function is based on the static one and is applied only to the above described cases for an empty set of bound variables.

**Proposition 1.** *Let $q$ be a query over an ontology $\mathcal{O}$, $s$ and $d$ the static and dynamic cost functions defined in Sections 5.1 and 5.2. The ordering induced by $s$ and $d$ is a total order over the axiom templates of $q$.*

*Proof.* The cost functions $s$ and $d$ are defined for all kinds of axiom templates and return two real numbers to each possible input. Since, according to Definition 10, the orders $\preceq_s$ and $\preceq_d$ are based on the addition of the two real numbers, addition of reals yields again a real number, and since $\leq$ is a total order over the reals, we immediately get that $\preceq_s$ and $\preceq_d$ are total orders. $\square$

It is obvious that the ordering of axiom templates does not affect soundness and completeness of a query evaluation algorithm.

## 6. Complex Axiom Template Optimizations

In this section, we first describe some optimizations that we have developed for complex axiom templates (Sections 6.1, 6.2) and then we present the procedure for evaluating queries (Section 6.3).

### 6.1 Axiom Template Rewriting

Some costly to evaluate axiom templates can be rewritten into axiom templates that can be evaluated more efficiently and yield an equivalent result. Before we go on to describe the axiom template rewriting technique, we define what a concept template is, which is useful throughout the section.

**Definition 12** (**Concept Template**). *Let $\mathcal{S}_q = (N_C, N_R, N_R, V_C, V_R, V_I)$ be a query signature w.r.t. a signature $\mathcal{S} = (N_C, N_R, N_I)$. A concept template over $\mathcal{S}_q$ is a $\mathcal{SROIQ}$ concept over $\mathcal{S}$, where one can also use concept variables from $V_C$ in place of concept names, role variables from $V_R$ in place of role names and individual variables from $V_I$ in place of individual names.*

**Definition 13 (Rewriting).** *Let $\mathcal{O}$ be an ontology, $at \in q$ an axiom template over $\mathcal{S}_q$, $t, t_1, \ldots t_n$ individuals or individual variables from $\mathcal{S}_q$, and $C, C_1, \ldots, C_n$ concept templates over $\mathcal{S}_q$. The function* rewrite *takes an axiom template and returns a set of axiom templates as follows:*

- *if $at = (C_1 \sqcap \ldots \sqcap C_n)(t)$, then* rewrite$(at) = \{C_1(t), \ldots, C_n(t)\}$*;*

- *if $at = C \sqsubseteq C_1 \sqcap \ldots \sqcap C_n$, then* rewrite$(at) = \{C \sqsubseteq C_1, \ldots, C \sqsubseteq C_n\}$*;*

- *if $at = C_1 \sqcup \ldots \sqcup C_n \sqsubseteq C$, then* rewrite$(at) = \{C_1 \sqsubseteq C, \ldots, C_n \sqsubseteq C\}$*;*

- *if $at = t_1 \approx \ldots \approx t_n$, then* rewrite$(at) = \{t_1 \approx t_2, t_2 \approx t_3, \ldots, t_{n-1} \approx t_n\}$*.*

To understand the intuition behind such transformation, we consider a query with only the axiom template: $?x \sqsubseteq \exists r.?y \sqcap A$. Its evaluation requires a quadratic number of consistency checks in the number of concepts (since $?x$ and $?y$ are concept variables). The rewriting yields: $?x \sqsubseteq A$ and $?x \sqsubseteq \exists r.?y$. The first axiom template is now evaluated with a cheap cache lookup (assuming that the concept hierarchy has been precomputed). For the second one, we only have to check the usually few resulting bindings for $?x$ combined with all other concept names for $?y$.

Note that Description Logics typically do not support n-ary equality axioms $t_1 \approx \ldots \approx t_n$, but only binary ones, whereas in OWL, one can typically also write n-ary equality axioms. Since our cost functions are only defined for binary equality axioms, we equivalently rewrite an n-ary one into several binary ones. One could even further optimize the evaluation of such atoms by just evaluating one binary equality axiom template and by then propagating the binding for the found equivalent individuals to the other equality axioms. This is valid since equality is a congruence relation.

## 6.2 Concept and Role Hierarchy Exploitation

The number of consistency checks required to evaluate a query can be further reduced by taking the concept and role hierarchies into account. Once the concepts and roles are classified (this can ideally be done before a system accepts queries), the hierarchies are stored in the reasoner's internal structures. We further use the hierarchies to prune the search space of solutions in the evaluation of certain axiom templates. We illustrate the intuition with the example Infection $\sqsubseteq \exists$hasCausalLinkTo.$?x$. If $A$ is not a solution and $B \sqsubseteq A$ holds, then $B$ is also not a solution. Thus, when searching for solutions for $?x$, we choose the next binding to test by traversing the concept hierarchy top-down. When we find a non-solution $A$, the subtree rooted in $A$ of the concept hierarchy can safely be pruned. Queries over ontologies with a large number of concepts and a deep concept hierarchy can, therefore, gain the maximum advantage from this optimization. We employ similar optimizations using the role hierarchies.

In the example above, we can prune the subconcepts of $A$ because $?x$ has positive polarity in the axiom template Infection $\sqsubseteq \exists$hasCausalLinkTo.$?x$., i.e., $?x$ occurs positively on the right hand side of the axiom template. In case a variable $?x$ has negative polarity in an axiom template of the form $C_1 \sqsubseteq C_2$, i.e., $?x$ occurs directly or indirectly under a negation on the right hand side of the axiom template or positively on the left-hand side of an axiom template, one can, instead, prune the superconcepts.

We next specify more precisely the polarity of a concept variable in a concept or axiom template.

**Definition 14** (**Concept Polarity**). *Let $?x \in V_C$ be a concept variable and $C, C_1, C_2, D$ concept templates, $r$ a role, and $n \in \mathbb{N}_0$. We define the polarity of $?x$ in $C$ as follows: $?x$ occurs positively in $?x$. Furthermore, $?x$ occurs positively (negatively)*

- *in $\neg D$ if $?x$ occurs negatively (positively) in $D$,*

- *in $C_1 \sqcap C_2$ or $C_1 \sqcup C_2$ if $?x$ occurs positively (negatively) in $C_1$ or $C_2$,*

- *in $\exists r.D$, $\forall r.D$, or $\geqslant n\ r.D$ if $?x$ occurs positively (negatively) in $D$,*

- *in $\leqslant n\ r.D$ if $?x$ occurs negatively (positively) in $D$*

- *in $= n\ r.D$ if $?x$ occurs in $D$.*

*We further say that $?x$ occurs positively (negatively) in $C_1 \sqsubseteq C_2$ if $?x$ occurs negatively (positively) in $C_1$ or positively (negatively) in $C_2$. Note that $?x$ can occur both positively and negatively in a concept template. We further define a partial function $\mathsf{pol}_c$ that maps a concept variable $?x$ and a concept template $C$ (axiom template of the form $C_1 \sqsubseteq C_2$) to $\mathsf{pos}$ if $?x$ occurs only positively in $C$ ($C_1 \sqsubseteq C_2$) and to $\mathsf{neg}$ if $?x$ occurs only negatively in $C$ ($C_1 \sqsubseteq C_2$).*

Note that no matter whether $?x$ occurs positively or negatively in a concept template $D$, in any concept template $C$ of the form $= n\ r.D$, $?x$ occurs positively as well as negatively. This is due to the fact that $C$ is equivalent to the concept template $\leqslant n\ r.D \sqcap \geqslant n\ r.D$ in which $?x$ occurs positively as well as negatively. Since the function $\mathsf{pol}_c$ is not defined for variables that appear both positively and negatively, the concept hierarchy cannot be exploited in this case. For example, consider the concept template $\neg ?x \sqcup \exists r.?x$, (axiom template $?x \sqsubseteq \exists r.?x$), where $?x$ appears negatively in $\neg ?x$ and positively in $\exists r.?x$. Now, let $\delta \in \Delta^{\mathcal{I}}$ be an arbitrary element from a model $\mathcal{I} = (\Delta^{\mathcal{I}}, \cdot^{\mathcal{I}})$ of the ontology. It is obvious that if $\delta$ is an instance of $\neg A \sqcup \exists r.A$ and either $A \sqsubseteq B$ or $B \sqsubseteq A$ holds, we cannot deduce that $\delta$ is an instance of $\neg B \sqcup \exists r.B$.

Before proving the correctness of the proposed optimization, we first show the relationship between entailment and concept membership, which is used in the subsequent proofs.

**Lemma 1.** *Let $q$ be a query over $\mathcal{O}$ w.r.t. the query signature $\mathcal{S}_q = (N_C, N_R, N_I, V_C, V_R, V_I)$, at $\in q$ be an axiom template of the form $C_1 \sqsubseteq C_2$ where $C_1$ and $C_2$ are concept templates and let $\mu$ be a mapping for at over $\mathcal{O}$. It holds that $\mathcal{O} \not\models \mu(C_1 \sqsubseteq C_2)$ iff there exists an interpretation $\mathcal{I} = (\Delta^{\mathcal{I}}, \cdot^{\mathcal{I}})$ and an element $\delta \in \Delta^{\mathcal{I}}$ such that $\mathcal{I} \models \mathcal{O}$ and $\delta \notin \mu(\neg C_1 \sqcup C_2)^{\mathcal{I}}$.*

*Proof.* $\mathcal{O} \not\models \mu(C_1 \sqsubseteq C_2)$ holds iff there exists an interpretation $\mathcal{I} = (\Delta^{\mathcal{I}}, \cdot^{\mathcal{I}})$ and an element $\delta \in \Delta^{\mathcal{I}}$ such that $\mathcal{I} \models \mathcal{O}$ and $\delta \in \mu(C_1)^{\mathcal{I}}$ and $\delta \notin \mu(C_2)^{\mathcal{I}}$, which holds iff $\delta \in \mu(C_1)^{\mathcal{I}}$ and $\delta \in \mu(\neg C_2)^{\mathcal{I}}$, which is equivalent to $\delta \in \mu(C_1 \sqcap \neg C_2)^{\mathcal{I}}$, which is equivalent to $\delta \in \mu(\neg(\neg C_1 \sqcup C_2))^{\mathcal{I}}$, which holds iff $\delta \notin \mu(\neg C_1 \sqcup C_2)^{\mathcal{I}}$. □

The following theorem holds for every axiom template of the form $C_1 \sqsubseteq C_2$. Note that we assume here that concept assertion templates of the form $C(a)$ are expressed as the equivalent axiom templates $\{a\} \sqsubseteq C$. We use $C_{\mu(?x)=A}$, where $A$ is a concept name, to denote the concept obtained by applying the extension of $\mu$ that also maps $?x$ to $A$.

**Theorem 1.** *Let $\mathcal{O}$ be an ontology, $A, B$ concept names such that $\mathcal{O} \models A \sqsubseteq B$, $C_1, C_2$ concept templates, $C_1 \sqsubseteq C_2$ an axiom template, $C = \neg C_1 \sqcup C_2$, $?x \in V_C$ a concept variable occurring in $C$ and $\mu$ a mapping that covers all variables of $C$ apart from $?x$.*

1. *For $\mathsf{pol}_c(?x, C) = \mathsf{pos}$ it holds that if $\mathcal{O} \not\models (C_1 \sqsubseteq C_2)_{\mu(?x)=B}$, then $\mathcal{O} \not\models (C_1 \sqsubseteq C_2)_{\mu(?x)=A}$.*

2. *For $\mathsf{pol}_c(?x, C) = \mathsf{neg}$ it holds that if $\mathcal{O} \not\models (C_1 \sqsubseteq C_2)_{\mu(?x)=A}$, then $\mathcal{O} \not\models (C_1 \sqsubseteq C_2)_{\mu(?x)=B}$.*

*Proof.* Due to Lemma 1, it suffices to show for some model $\mathcal{I} = (\Delta^{\mathcal{I}}, \cdot^{\mathcal{I}})$ of $\mathcal{O}$ and some element $\delta \in \Delta^{\mathcal{I}}$ the following (which is formalized in contrapositive form):

1. For $\mathsf{pol}_c(?x, C) = \mathsf{pos}$ it holds that if $\delta \in (C_{\mu(?x)=A})^{\mathcal{I}}$, then $\delta \in (C_{\mu(?x)=B})^{\mathcal{I}}$.

2. For $\mathsf{pol}_c(?x, C) = \mathsf{neg}$ it holds that if $\delta \in (C_{\mu(?x)=B})^{\mathcal{I}}$, then $\delta \in (C_{\mu(?x)=A})^{\mathcal{I}}$.

We prove the claim by induction on the structure of the concept template $C$:

- For $C = ?x$, $?x$ occurs positively in $C$. Now, if $\delta \in (?x_{\mu(?x)=A})^{\mathcal{I}}$, that is $\delta \in A^{\mathcal{I}}$, it is easy to see that $\delta \in B^{\mathcal{I}}$ since $\mathcal{O} \models A \sqsubseteq B$ by assumption. Hence, $\delta \in (?x_{\mu(?x)=B})^{\mathcal{I}}$.

- For $C = \neg D$ and $\mathsf{pol}_c(?x, C) = \mathsf{pos}$, if $\delta \in (\neg D_{\mu(?x)=A})^{\mathcal{I}}$, we have to show that $\delta \in (\neg D_{\mu(?x)=B})^{\mathcal{I}}$. Note that $\mathsf{pol}_c(?x, D) = \mathsf{neg}$. In contrary to what is to be shown, assume that $\delta \in (D_{\mu(?x)=B})^{\mathcal{I}}$. Since $\mathcal{O} \models A \sqsubseteq B$ and by induction hypothesis $\delta \in (D_{\mu(?x)=A})^{\mathcal{I}}$ and $\delta \in (\neg D_{\mu(?x)=A})^{\mathcal{I}}$ which is a contradiction. The proof is analogous for $\mathsf{pol}_c(?x, C) = \mathsf{neg}$.

- For $C = C_1 \sqcap C_2$ and $\mathsf{pol}_c(?x, C) = \mathsf{pos}$, if $\delta \in ((C_1 \sqcap C_2)_{\mu(?x)=A})^{\mathcal{I}}$, then $\delta \in (C_{1\mu(?x)=A})^{\mathcal{I}}$ and $\delta \in (C_{2\mu(?x)=A})^{\mathcal{I}}$. Since $\mathcal{O} \models A \sqsubseteq B$ and by induction hypothesis, $\delta \in (C_{1\mu(?x)=B})^{\mathcal{I}}$ and $\delta \in (C_{2\mu(?x)=B})^{\mathcal{I}}$. Thus, $\delta \in ((C_1 \sqcap C_2)_{\mu(?x)=B})^{\mathcal{I}}$. The proof is analogous for $\mathsf{pol}_c(?x, C) = \mathsf{neg}$.

- The proof for $C_1 \sqcup C_2$ is analogous to the one for $C_1 \sqcap C_2$.

- For $C = \exists r.D$ and $\mathsf{pol}_c(?x, C) = \mathsf{pos}$, if $\delta \in ((\exists r.D)_{\mu(?x)=A})^{\mathcal{I}}$, then $\delta$ has at least one $r$-successor, say $\delta'$, that is an instance of $D_{\mu(?x)=A}$. Since $\mathcal{O} \models A \sqsubseteq B$ and by induction hypothesis, $\delta' \in D_{\mu(?x)=B}$. Hence, $\delta \in (\exists r.(D_{\mu(?x)=B}))^{\mathcal{I}} = ((\exists r.D)_{\mu(?x)=B})^{\mathcal{I}}$. The proof is analogous for $\mathsf{pol}_c(?x, C) = \mathsf{neg}$.

- For $C = \forall r.D$ and $\mathsf{pol}_c(?x, C) = \mathsf{pos}$, if $\delta \in ((\forall r.D)_{\mu(?x)=A})^{\mathcal{I}}$, then $\delta \in (\forall r.(D)_{\mu(?x)=A})^{\mathcal{I}}$ and each $r$-successors of $\delta$ is an instance of $D_{\mu(?x)=A}$. Since $\mathcal{O} \models A \sqsubseteq B$ and by induction hypothesis, these $r$-successors are also instances of $D_{\mu(?x)=B}$. Hence, $\delta \in (\forall r.(D_{\mu(?x)=B}))^{\mathcal{I}} = ((\forall r.D)_{\mu(?x)=B})^{\mathcal{I}}$. The proof is analogous for $\mathsf{pol}_c(?x, C) = \mathsf{neg}$.

- For $C = \geqslant n\ r.D$ and $\mathsf{pol}_c(?x, C) = \mathsf{pos}$, if $\delta \in ((\geqslant n\ r.D)_{\mu(?x)=A})^{\mathcal{I}}$, then $\delta$ has at least $n$ distinct $r$-successors which are instances of $D_{\mu(?x)=A}$. Since $\mathcal{O} \models A \sqsubseteq B$ and by induction hypothesis, these successors are instances of $D_{\mu(?x)=B}$. Hence, $\delta$ has at least $n$ distinct $r$-successors that are instances of $D_{\mu(?x)=B}$ and, therefore, $\delta \in (\geqslant n\ r.(D)_{\mu(?x)=B})^{\mathcal{I}} = ((\geqslant n\ r.D)_{\mu(?x)=B})^{\mathcal{I}}$. The proof is analogous for $\mathsf{pol}_c(?x, C) = \mathsf{neg}$.

- For $C = \leqslant n\ r.D$ and $\mathsf{pol}_c(?x, C) = \mathsf{pos}$, if $\delta \in ((\leqslant n\ r.D)_{\mu(?x)=A})^{\mathcal{I}}$, we have to show that $\delta \in ((\leqslant n\ r.D)_{\mu(?x)=B})^{\mathcal{I}}$. Note that $\mathsf{pol}_c(?x, D) = \mathsf{neg}$. In contrary to what is to be shown, assume that $\delta \in (\neg(\leqslant n\ r.D)_{\mu(?x)=B})^{\mathcal{I}}$, i.e., $\delta \in ((\geqslant n+1\ r.D)_{\mu(?x)=B})^{\mathcal{I}}$. Hence, $\delta$ has at least $n+1$ distinct $r$-successors which are instances of $D_{\mu(?x)=B}$. Since $\mathsf{pol}_c(?x, D) = \mathsf{neg}$ and by induction hypothesis, these $D_{\mu(?x)=B}$ instances are also $D_{\mu(?x)=A}$ instances and $\delta \in (\geqslant n+1\ r.(D)_{\mu(?x)=A})^{\mathcal{I}} = ((\geqslant n+1\ r.D)_{\mu(?x)=A})^{\mathcal{I}}$, which is a contradiction. The proof is analogous for $\mathsf{pol}_c(?x, C) = \mathsf{neg}$.

- For $C = (= n\ r.D)$, the polarity of $?x$ in $C$ is always positive and negative, so $\mathsf{pol}_c(?x, C)$ is undefined and the case cannot occur. □

We now extend this optimization to the case of role variables and we first define the polarity of a role variable in a concept or axiom template.

**Definition 15** (**Role Polarity**). *Let* $?x \in V_R$ *be a role variable,* $C, C_1, C_2, D$ *concept templates,* $r$ *a role, and* $n \in \mathbb{N}_0$. *We define the polarity of* $?x$ *in* $C$ *as follows:* $?x$ *occurs positively in* $\exists ?x.D$, $\exists ?x^-.D$, $\geqslant n\ ?x.D$, $\geqslant n\ ?x^-.D$, $= n\ ?x^-.D$; $?x$ *occurs negatively in* $\forall ?x.D$, $\forall ?x^-.D$, $\leqslant n\ ?x.D$, $\leqslant n\ ?x^-.D$, $= n\ ?x.D$, *and* $= n\ ?x^-.D$. *Furthermore,* $?x$ *occurs positively (negatively)*

- *in* $\neg D$ *if* $?x$ *occurs negatively (positively) in* $D$,

- *in* $C_1 \sqcap C_2$ *or* $C_1 \sqcup C_2$ *if* $?x$ *occurs positively (negatively) in* $C_1$ *or* $C_2$,

- *in* $\exists r.D$, $\exists ?x.D$, $\exists ?x^-.D$, $\geqslant n\ r.D$, $\geqslant n\ ?x.D$, $\geqslant n\ ?x^-.D$, $\forall r.D$, $\forall ?x.D$, *or* $\forall ?x^-.D$ *if* $?x$ *occurs positively (negatively) in* $D$,

- *in* $\leqslant n\ r.D$, $\leqslant n\ ?x.D$, *or* $\leqslant n\ ?x^-.D$ *if* $?x$ *occurs negatively (positively) in* $D$,

- *in* $= n\ r.D$ *if* $?x$ *occurs in* $D$.

*We further say that* $?x$ *occurs positively (negatively) in* $C_1 \sqsubseteq C_2$ *if* $?x$ *occurs negatively (positively) in* $C_1$ *or positively (negatively) in* $C_2$. *We define a partial function* $\mathsf{pol}_r$ *that maps a role variable* $?x$ *and a concept template* $C$ *(axiom template of the form* $C_1 \sqsubseteq C_2$*) to* $\mathsf{pos}$ *if* $?x$ *occurs only positively in* $C$ *(*$C_1 \sqsubseteq C_2$*) and to* $\mathsf{neg}$ *if* $?x$ *occurs only negatively in* $C$ *(*$C_1 \sqsubseteq C_2$*).*

Note also that we do not make any assumption about occurrences of $?x$ in $D$ in the first part of the definition.

We now show, that the hierarchy optimization is also applicable to role variables, provided they occur only positively or only negatively.

**Theorem 2.** *Let* $\mathcal{O}$ *be an ontology,* $r, s$ *role names such that* $\mathcal{O} \models r \sqsubseteq s$, $C_1, C_2$ *concept templates,* $C_1 \sqsubseteq C_2$ *an axiom template,* $C = \neg C_1 \sqcup C_2$, $?x \in V_R$ *a role variable occurring in* $C$ *and* $\mu$ *a mapping that covers all variables of* $C$ *apart from* $?x$.

1. *For* $\mathsf{pol}_r(?x, C) = \mathsf{pos}$ *it holds that if* $\mathcal{O} \not\models (C_1 \sqsubseteq C_2)_{\mu(?x)=s}$, *then* $\mathcal{O} \not\models (C_1 \sqsubseteq C_2)_{\mu(?x)=r}$.

2. *For* $\mathsf{pol}_r(?x, C) = \mathsf{neg}$ *it holds that if* $\mathcal{O} \not\models (C_1 \sqsubseteq C_2)_{\mu(?x)=r}$, *then* $\mathcal{O} \not\models (C_1 \sqsubseteq C_2)_{\mu(?x)=s}$.

*Proof.* Due to Lemma 1, it suffices to show for some model $\mathcal{I} = (\Delta^{\mathcal{I}}, \cdot^{\mathcal{I}})$ of $\mathcal{O}$ and some element $\delta \in \Delta^{\mathcal{I}}$ the following (which is formalized in contrapositive form):

1. For $\mathsf{pol}_r(?x, C) = \mathsf{pos}$ it holds that if $\delta \in (C_{\mu(?x)=r})^{\mathcal{I}}$, then $\delta \in (C_{\mu(?x)=s})^{\mathcal{I}}$.

2. For $\mathsf{pol}_r(?x, C) = \mathsf{neg}$ it holds that if $\delta \in (C_{\mu(?x)=s})^{\mathcal{I}}$, then $\delta \in (C_{\mu(?x)=r})^{\mathcal{I}}$.

We prove the claim by induction on the structure of the concept template $C$:

- For $C = \exists ?x.D$, where $D$ is a concept template that does not contain $?x$. We have $\mathsf{pol}_r(?x, C) = \mathsf{pos}$. Assume, $\delta \in ((\exists ?x.D)_{\mu(?x)=r})^{\mathcal{I}}$, that is, $\delta \in (\exists r.\mu(D))^{\mathcal{I}}$. Then there is some $\delta' \in \Delta^{\mathcal{I}}$ such that $\langle \delta, \delta' \rangle \in r^{\mathcal{I}}$ and $\delta' \in \mu(D)^{\mathcal{I}}$. Since $\mathcal{O} \models r \sqsubseteq s$, we also have $\langle \delta, \delta' \rangle \in s^{\mathcal{I}}$ and, therefore, $\delta \in (\exists s.\mu(D))^{\mathcal{I}} = ((\exists ?x.D)_{\mu(?x)=s})^{\mathcal{I}}$.

- For $C = \forall ?x.D$, where $D$ is a concept template that does not contain $?x$. We have $\mathsf{pol}_r(?x, C) = \mathsf{neg}$. If $\delta \in ((\forall ?x.D)_{\mu(?x)=s})^{\mathcal{I}}$, we have to show that $\delta \in ((\forall ?x.D)_{\mu(?x)=r})^{\mathcal{I}}$. In contrary to what is to be shown, assume that $\delta \in (\neg(\forall ?x.D)_{\mu(?x)=r})^{\mathcal{I}}$, i.e., $\delta \in (\exists r.\mu(\neg D))^{\mathcal{I}}$. Hence, there is some $\delta' \in \Delta^{\mathcal{I}}$ such that $\langle \delta, \delta' \rangle \in r^{\mathcal{I}}$ and $\delta' \in \mu(\neg D)^{\mathcal{I}}$. Since $\mathcal{O} \models r \sqsubseteq s$, we also have $\langle \delta, \delta' \rangle \in s^{\mathcal{I}}$ and, therefore, $\delta \notin (\forall s.\mu(D))^{\mathcal{I}} = ((\forall ?x.D)_{\mu(?x)=s})^{\mathcal{I}}$, which is a contradiction.

- For $C = \geqslant n \ ?x.D$ where $D$ is a concept template that does not contain $?x$. We have $\mathsf{pol}_r(?x, C) = \mathsf{pos}$. Assume, $\delta \in ((\geqslant n \ ?x.D)_{\mu(?x)=r})^{\mathcal{I}}$, that is $\delta \in (\geqslant n \ r.\mu(D))^{\mathcal{I}}$ and $\delta$ has at least $n$ distinct $r$-successors which are instances of $\mu(D)$. Since $\mathcal{O} \models r \sqsubseteq s$ these $r$-successors are also $s$-successors of $\delta$ and, therefore, $\delta \in (\geqslant n \ s.\mu(D))^{\mathcal{I}} = ((\geqslant n \ ?x.D)_{\mu(?x)=s})^{\mathcal{I}}$.

- For $C = \leqslant n \ ?x.D$ where $C$ is a concept template that does not contain $?x$. We have $\mathsf{pol}_r(?x, C) = \mathsf{neg}$. If $\delta \in ((\leqslant n \ ?x.D)_{\mu(?x)=s})^{\mathcal{I}}$, we have to show that $\delta \in ((\leqslant n \ ?x.D)_{\mu(?x)=r})^{\mathcal{I}}$. In contrary to what is to be shown, assume that $\delta \in (\neg(\leqslant n \ ?x.D)_{\mu(?x)=r})^{\mathcal{I}}$, i.e., $\delta \in (\geqslant n+1 \ r.\mu(D))^{\mathcal{I}}$. Hence, $\delta$ has at least $n+1$ distinct $r$-successors, which are instances of $\mu(D)$. Since $\mathcal{O} \models r \sqsubseteq s$, these $r$-successors are also $s$-successors and $\delta \in ((\geqslant n+1 \ s.\mu(D)))^{\mathcal{I}} = ((\geqslant n+1 \ ?x.D)_{\mu(?x)=s})^{\mathcal{I}}$, which is a contradiction.

- For $C = C_1 \sqcap C_2$ and $\mathsf{pol}_r(?x, C) = \mathsf{pos}$, if $\delta \in ((C_1 \sqcap C_2)_{\mu(?x)=r})^{\mathcal{I}}$, then $\delta \in (C_{1\mu(?x)=r})^{\mathcal{I}}$ and $\delta \in (C_{2\mu(?x)=r})^{\mathcal{I}}$. Since $\mathcal{O} \models r \sqsubseteq s$ and by the induction hypothesis, $\delta \in (C_{1\mu(?x)=s})^{\mathcal{I}}$ and $\delta \in (C_{2\mu(?x)=s})^{\mathcal{I}}$. Thus, $\delta \in ((C_1 \sqcap C_2)_{\mu(?x)=s})^{\mathcal{I}}$. The proof is analogous for $\mathsf{pol}_r(?x, C) = \mathsf{neg}$.

- The proof for $C_1 \sqcup C_2$ is analogous to the one for $C_1 \sqcap C_2$.

- For $C = \neg D$ and $\mathsf{pol}_r(?x, C) = \mathsf{pos}$, if $\delta \in (\neg D_{\mu(?x)=r})^{\mathcal{I}}$, we have to show that $\delta \in (\neg D_{\mu(?x)=s})^{\mathcal{I}}$. Note that $\mathsf{pol}_r(?x, D) = \mathsf{neg}$. In contrary to what is to be shown, assume that $\delta \in (D_{\mu(?x)=s})^{\mathcal{I}}$. Since $\mathcal{O} \models r \sqsubseteq s$ and by induction hypothesis $\delta \in (D_{\mu(?x)=r})^{\mathcal{I}}$ and $\delta \in (\neg D_{\mu(?x)=r})^{\mathcal{I}}$ which is a contradiction. The proof is analogous for $\mathsf{pol}_r(?x, C) = \mathsf{neg}$.

- For $C = \exists p.D$ and $\mathsf{pol}_r(?x, C) = \mathsf{pos}$, we also have $\mathsf{pol}_r(?x, D) = \mathsf{pos}$. Now, if $\delta \in ((\exists p.D)_{\mu(?x)=r})^{\mathcal{I}}$, then $\delta$ has at least one $p$-successor that is an instance of $D_{\mu(?x)=r}$. Since $\mathcal{O} \models r \sqsubseteq s$ and by induction hypothesis, this $p$-successor is an instance of $D_{\mu(?x)=s}$. Hence, $\delta \in ((\exists p.D)_{\mu(?x)=s})^{\mathcal{I}}$. The proof is analogous for $\mathsf{pol}_r(?x, C) = \mathsf{neg}$.

- For $C = \exists ?x.D$ and $\mathsf{pol}_r(?x, C) = \mathsf{pos}$, we also have $\mathsf{pol}_r(?x, D) = \mathsf{pos}$. Note that $?x$ occurs in $D$ since otherwise the case is handled already above. Now, if $\delta \in ((\exists ?x.D)_{\mu(?x)=r})^{\mathcal{I}}$, then $\delta$ has at least one $r$-successor which is an instance of $D_{\mu(?x)=r}$. Since $\mathcal{O} \models r \sqsubseteq s$ and by induction hypothesis, $\delta$ has at least one $s$-successor that is an instance of $D_{\mu(?x)=s}$. Hence, $\delta \in ((\exists ?x.D)_{\mu(?x)=s})^{\mathcal{I}}$.

- For $C = \forall p.D$ and $\mathsf{pol}_r(?x, C) = \mathsf{pos}$, we also have $\mathsf{pol}_r(?x, D) = \mathsf{pos}$. Now, if $\delta \in ((\forall p.D)_{\mu(?x)=r})^{\mathcal{I}}$, then $\delta \in (\forall p.(D)_{\mu(?x)=r})^{\mathcal{I}}$ and each $p$-successor of $\delta$ is an instance of $D_{\mu(?x)=r}$. Since $\mathcal{O} \models r \sqsubseteq s$ and by induction hypothesis, these $p$-successors are also instances of $D_{\mu(?x)=s}$. Hence, $\delta \in (\forall p.(D_{\mu(?x)=s}))^{\mathcal{I}} = ((\forall p.D)_{\mu(?x)=s})^{\mathcal{I}}$. The proof is analogous for $\mathsf{pol}_r(?x, C) = \mathsf{neg}$.

- For $C = \forall ?x.D$ and $\mathsf{pol}_r(?x, C) = \mathsf{neg}$, we also have $\mathsf{pol}_r(?x, D) = \mathsf{neg}$. Note that $?x$ occurs in $D$ since otherwise the case is handled already above. Now, if $\delta \in ((\forall ?x.D)_{\mu(?x)=s})^{\mathcal{I}}$, we have to show that $\delta \in ((\forall ?x.D)_{\mu(?x)=r})^{\mathcal{I}}$. In contrary to what is to be shown, assume that $\delta \notin ((\forall ?x.D)_{\mu(?x)=r})^{\mathcal{I}}$, i.e., $\delta \in (\exists r.(\neg D)_{\mu(?x)=r})^{\mathcal{I}}$. Hence, there is some $\delta' \in \Delta^{\mathcal{I}}$ such that $\langle \delta, \delta' \rangle \in r^{\mathcal{I}}$ and $\delta' \in ((\neg D)_{\mu(?x)=r})^{\mathcal{I}}$. Since $\mathcal{O} \models r \sqsubseteq s$, $\delta'$ is also an $s$-successor of $\delta$ and, by induction hypothesis, we have $\delta' \in ((\neg D)_{\mu(?x)=s})^{\mathcal{I}}$ which is a contradiction.

- For $C = \geqslant n\ p.D$ and $\mathsf{pol}_r(?x, C) = \mathsf{pos}$, if $\delta \in ((\geqslant n\ p.D)_{\mu(?x)=r})^{\mathcal{I}}$, then $\delta$ has at least $n$ distinct $p$-successors that are instances of $D_{\mu(?x)=r}$. Since $\mathcal{O} \models r \sqsubseteq s$ and by induction hypothesis, these $p$-successors are also instances of $D_{\mu(?x)=s}$. Hence, $\delta \in ((\geqslant n\ p.D)_{\mu(?x)=s})^{\mathcal{I}}$. The proof is analogous for $\mathsf{pol}_r(?x, C) = \mathsf{neg}$

- For $C = \geqslant n\ ?x.D$ and $\mathsf{pol}_r(?x, C) = \mathsf{pos}$, we also have $\mathsf{pol}_r(?x, D) = \mathsf{pos}$. Note that $?x$ occurs in $D$ since otherwise the case is handled already above. Now, if $\delta \in ((\geqslant n\ ?x.D)_{\mu(?x)=r})^{\mathcal{I}}$, then $\delta$ has at least $n$ distinct $r$-successors which are instances of $D_{\mu(?x)=r}$. Since $\mathcal{O} \models r \sqsubseteq s$ and by induction hypothesis, $\delta$ has at least $n$ distinct $s$-successors that are instances of $D_{\mu(?x)=s}$. Hence, $\delta \in ((\geqslant n\ ?x.D)_{\mu(?x)=s})^{\mathcal{I}}$.

- For $C = \leqslant n\ p.D$ and $\mathsf{pol}_r(?x, C) = \mathsf{pos}$, if $\delta \in ((\leqslant n\ p.D)_{\mu(?x)=r})^{\mathcal{I}}$, we have to show that $\delta \in ((\leqslant n\ p.D)_{\mu(?x)=s})^{\mathcal{I}}$. Note that $\mathsf{pol}_r(?x, D) = \mathsf{neg}$. In contrary to what is to be shown, assume that $\delta \in (\neg(\leqslant n\ p.D)_{\mu(?x)=s})^{\mathcal{I}}$, i.e., $\delta \in ((\geqslant n+1\ p.D)_{\mu(?x)=s})^{\mathcal{I}}$. Hence, $\delta$ has at least $n+1$ distinct $p$-successors which are instances of $D_{\mu(?x)=s}$. Since $\mathsf{pol}_r(?x, D) = \mathsf{neg}$ and by induction hypothesis, these $D_{\mu(?x)=s}$ instances are also $D_{\mu(?x)=r}$ instances and $\delta \in (\geqslant n+1\ p.(D)_{\mu(?x)=r})^{\mathcal{I}} = ((\geqslant n+1\ p.D)_{\mu(?x)=r})^{\mathcal{I}}$, which is a contradiction. The proof is analogous for $\mathsf{pol}_r(?x, C) = \mathsf{neg}$.

- For $C = \leqslant n\ ?x.D$ and $\mathsf{pol}_r(?x, D) = \mathsf{pos}$, we have $\mathsf{pol}_r(?x, D) = \mathsf{pos}$. Note that $?x$ occurs in $D$ since otherwise the case is handled already above. If $\delta \in ((\leqslant n\ ?x.D)_{\mu(?x)=s})^{\mathcal{I}}$ we have to show that $\delta \in ((\leqslant n\ ?x.D)_{\mu(?x)=r})^{\mathcal{I}}$. In contrary

---

**Algorithm 2** getPossibleMappings($\mathcal{O}, ?x, at, \mu$)

---

**Input:**   $\mathcal{O}$: the queried $\mathcal{SROIQ}$ ontology

        $?x$: a concept or role variable

        $at$: an axiom template in which $?x$ occurs

        $\mu$: a mapping with $?x \in \mathrm{dom}(\mu)$

**Output:** a set of mappings

 1:   $S := \emptyset$

 2:   **if** $?x \in V_C$ **then**

 3:     **if** $\mathrm{pol}_c(?x, at) = \mathsf{pos}$ **then**

 4:       $S := \{\mu' \mid \mu'(?x) = A, A$ is a direct subconcept of $\mu(?x)$ in $\mathcal{O}$,

              $\mu'(?y) = \mu(?y)$ for $?y \in \mathrm{dom}(\mu) \setminus \{?x\}\}$

 5:     **else**

 6:       $S := \{\mu' \mid \mu'(?x) = A, A$ is a direct superconcept of $\mu(?x)$ in $\mathcal{O}$,

              $\mu'(?y) = \mu(?y)$ for $?y \in \mathrm{dom}(\mu) \setminus \{?x\}\}$

 7:     **end if**

 8:   **else**

 9:     **if** $\mathrm{pol}_r(?x, at) = \mathsf{pos}$ **then**

10:       $S := \{\mu' \mid \mu'(?x) = r, r$ is a direct subrole of $\mu(?x)$ in $\mathcal{O}$,

              $\mu'(?y) = \mu(?y)$ for $?y \in \mathrm{dom}(\mu) \setminus \{?x\}\}$

11:     **else**

12:       $S := \{\mu' \mid \mu'(?x) = r, r$ is a direct superrole of $\mu(?x)$ in $\mathcal{O}$,

              $\mu'(?y) = \mu(?y)$ for $?y \in \mathrm{dom}(\mu) \setminus \{?x\}\}$

13:     **end if**

14:   **end if**

15:   **return** $S$

---

to what is to be shown, assume that $\delta \in (\neg(\leqslant n\ ?x.D)_{\mu(?x)=r})^{\mathcal{I}}$, i.e., $\delta \in ((\geqslant n+1\ ?x.D)_{\mu(?x)=r})^{\mathcal{I}}$. Hence, $\delta$ has at least $n+1$ distinct $r$-successors which are instances of $D_{\mu(?x)=r}$. Since $\mathcal{O} \models r \sqsubseteq s$, and by induction hypothesis, these $r$-successors are also $s$-successors and instances of $D_{\mu(?x)=s}$. Hence, $\delta \in ((\geqslant n+1\ ?x.D)_{\mu(?x)=s})^{\mathcal{I}}$ and $\delta \in ((\leqslant n\ ?x.D)_{\mu(?x)=s})^{\mathcal{I}}$, which is a contradiction.

- For $C = (= n\ ?x.D)$ or $C = (= n\ r.D)$, the polarity of $?x$ in $C$ is always positive and negative, so $\mathrm{pol}_r(?x, C)$ is undefined and the case cannot occur.

- The cases for $?x$ occurring in the form of an inverse ($?x^-$) are analogous, given that $\mathcal{O} \models r \sqsubseteq s$ iff $\mathcal{O} \models r^- \sqsubseteq s^-$.     $\square$

Algorithm 2, which we explain in detail in Section 6.3, shows how we use the above theorems to create possible concept and role mappings for a concept or role variable $?x$ that appears only positively or only negatively in an axiom template $C_1 \sqsubseteq C_2$.

## 6.3 Query Answering Algorithm

Algorithm 3 shows an optimized way of evaluating queries using static ordering. First, axiom templates are simplified where possible (method **rewrite** in line 1). Next, the method

---

**Algorithm 3** evaluate$(\mathcal{O}, q)$

---

**Input:** $\mathcal{O}$: the queried $\mathcal{SROIQ}$ ontology

        $q$: a query over $\mathcal{O}$

**Output:** a set of solutions for evaluating $q$ over $\mathcal{O}$

 1: $At := \mathsf{rewrite}(q)$

 2: $At^1, \ldots, At^m := \mathsf{connectedComponents}(At)$

 3: **for** j=1, ..., m **do**

 4:     $R_j := \{\mu_0 \mid \mathrm{dom}(\mu_0) = \emptyset\}$

 5:     $at_1, \ldots, at_n := \mathsf{order}(At^j)$

 6:     **for** $i = 1, \ldots, n$ **do**

 7:        $R := \emptyset$

 8:        **for each** $\mu \in R_j$ **do**

 9:           **if** $\mathsf{isSimple}(at_i)$ **and** $\mathrm{Vars}(at_i) \setminus \mathrm{dom}(\mu) \neq \emptyset$ **then**

10:             $R := R \cup \{\mu' \cup \mu \mid \mu' \in \mathsf{callSpecificReasonerTask}(\mu(at_i))\}$

11:           **else if** $\mathrm{Vars}(at_i) \setminus \mathrm{dom}(\mu) = \emptyset$ **then**

12:             **if** $\mathcal{O} \models \mu(at_i)$ **then**

13:                $R := R \cup \{\mu\}$

14:             **end if**

15:           **else**

16:             $V_{\mathrm{opt}} := \{?x \mid ?x \notin \mathrm{dom}(\mu), \text{ Theorem 1 or 2 applies to } ?x \text{ and } at_i\}$

17:             $B := \mathsf{initializeVariableMappings}(\mathcal{O}, at_i, \mu, V_{\mathrm{opt}})$

18:             **while** $B \neq \emptyset$ **do**

19:                $\mu' := \mathsf{removeMapping}(B)$

20:                **if** $\mathcal{O} \models \mu'(at_i)$ **then**

21:                   $R := R \cup \{\mu'' \mid \mu''(?x) = \mu'(?x) \text{ if } ?x \notin V_{\mathrm{opt}} \text{ and}$

                                 $\mu''(?x) = C \text{ if } ?x \in V_{\mathrm{opt}} \cap V_C, \mathcal{O} \models C \equiv \mu'(?x) \text{ and}$

                                 $\mu''(?x) = r \text{ if } ?x \in V_{\mathrm{opt}} \cap V_R, \mathcal{O} \models r \equiv \mu'(?x)\}$

22:                  **for each** $?x \in V_{\mathrm{opt}}$ **do**

23:                     $B := B \cup \mathsf{getPossibleMappings}(\mathcal{O}, ?x, at_i, \mu')$

24:                  **end for**

25:                **end if**

26:             **end while**

27:           **end if**

28:        **end for**

29:        $R_j := R$

30:     **end for**

31: **end for**

32: $R_{\mathrm{ans}} := \{\mu_1 \cup \ldots \cup \mu_m \mid \mu_j \in R_j, 1 \leq j \leq m\}$

33: **return** $R_{\mathrm{ans}}$

---

$\mathsf{connectedComponents}$ (line 2) partitions the axiom templates into sets of connected components, i.e., within a component the templates share common variables, whereas between components there are no shared variables. Unconnected components unnecessarily increase the amount of intermediate results and, instead, one can simply combine the results for the

---

**Algorithm 4** initializeVariableMappings($\mathcal{O}, at, \mu, V_{\text{opt}}$)

---

**Input:** $\mathcal{O}$: the queried $\mathcal{SROIQ}$ ontology

$\qquad$ $at$: an axiom template

$\qquad$ $\mu$: a partial mapping

$\qquad$ $V_{\text{opt}}$: the variables of $at$ to which Theorem 1 or 2 applies

**Output:** a set of mappings

1: $S := \{\mu\}$
2: **for each** $?x \in \text{Vars}(at) \setminus \text{dom}(\mu)$ **do**
3: $\quad$ $R := \emptyset$
4: $\quad$ **if** $?x \in V_C$ and $?x \in V_{\text{opt}}$ **then**
5: $\quad\quad$ **for each** $\mu' \in S$ **do**
6: $\quad\quad\quad$ **if** $\text{pol}_c(?x, at) = \text{pos}$ **then**
7: $\quad\quad\quad\quad$ $\mu'(?x) := \top$
8: $\quad\quad\quad$ **else**
9: $\quad\quad\quad\quad$ $\mu'(?x) := \bot$
10: $\quad\quad\quad$ **end if**
11: $\quad\quad\quad$ $R := R \cup \{\mu'\}$
12: $\quad\quad$ **end for**
13: $\quad$ **else if** $?x \in V_R$ and $?x \in V_{\text{opt}}$ **then**
14: $\quad\quad$ **for each** $\mu' \in S$ **do**
15: $\quad\quad\quad$ **if** $\text{pol}_r(?x, at) = \text{pos}$ **then**
16: $\quad\quad\quad\quad$ $\mu'(?x) := \top_r$
17: $\quad\quad\quad$ **else**
18: $\quad\quad\quad\quad$ $\mu'(?x) := \bot_r$
19: $\quad\quad\quad$ **end if**
20: $\quad\quad\quad$ $R := R \cup \{\mu'\}$
21: $\quad\quad$ **end for**
22: $\quad$ **else**
23: $\quad\quad$ $R := \{\mu' \mid \mu'(?x) = a, a \in N_C^{\mathcal{O}} \text{ or } a \in N_R^{\mathcal{O}} \text{ or } a \in N_I^{\mathcal{O}} \text{ and } \mu'(?y) = \mu_1(?y)$
$\qquad\qquad\qquad$ for $\mu_1 \in S$ and $?y \in \text{dom}(\mu_1)\}$
24: $\quad$ **end if**
25: $\quad$ $S := R$
26: **end for**
27: **return** S

---

components in the end (line 32). For each component, we proceed as described below: we first determine an order (method **order** in line 5) as described in Section 5. For a simple axiom template, which contains so far unbound variables, we call a specialized reasoner method to retrieve entailed results, i.e., mappings for unbound variables (**callSpecificReasonerTask** in line 10). Note that the mappings $\mu'$ do not assign values to any of the variables covered by the already computed (partial) solution $\mu$ since we instantiate the atom $at_i$ by $\mu$. This allows for defining the union of $\mu$ and $\mu'$ by setting $(\mu \cup \mu')(v) = \mu(v)$ if $v \in \text{dom}(\mu)$, and $(\mu \cup \mu')(v) = \mu'(v)$ otherwise. For templates with all their variables bound, we check whether the mappings lead to entailed axioms (lines 11 to 14). For all other cases, i.e., for complex axiom templates with unbound variables, we check which compatible mappings yield an entailed axiom (lines 15 to 27). In particular, we first initialize a set $B$ of candidate mappings for the unbound variables of the axiom template (line 17, which refers to Algorithm 4). Algorithm 4 initializes the unbound variables of axiom templates on which Theorem 1 or 2 applies to $\top$ ($\top_r$) or $\bot$ ($\bot_r$) depending on whether the respective polarity function returns **pos** or **neg**. For template variables on which the optimization is not applicable, all compatible mappings are returned. The method **removeMapping** (line 19) returns a mapping from $B$ and deletes this mapping from $B$. We then instantiate the axiom template and check entailment. In case the entailment holds, we first extend the set $R$ with the current mapping $\mu'$ and with mappings that map the optimization variables to equivalent concepts or roles of the respective variable mappings in $\mu'$ (line 21) and we afterwards extend the set $B$ of possible mappings for the variables to which the hierarchy optimization is applicable (**getPossibleMappings** in line 23). For example, if we just checked a mapping $\mu$ that maps a concept variable $?x$ to the concept $A$ and $?x$ only occurs positively in the axiom template, then we add to the set $B$ all mappings that map $?x$ to a direct subconcept[3] of $A$ (see Algorithm 2 line 4). In the implementation we use a more involved procedure, i.e., in order to avoid checking entailment of an instantiated axiom template more than once with the same mapping, which can be the case with the concept (role) hierarchy traversal that we perform, we keep track of already processed mappings and check only those that have not been checked in a previous iteration of the while loop (lines 18 to 26). For ease of presentation, this is not shown in Algorithm 3. We then repeat the procedure until $B$ is empty (lines 18 to 26).

For the dynamic ordering, Algorithm 3 has to be changed as follows: We first compute the number of axiom templates in $At^j$; $n := |At^j|$. We then swap line 5 and line 6, i.e., instead of ordering all axiom templates before the loop that evaluates the axiom templates, we order within the for loop. The function **order** gets as additional input parameter the set of currently computed solutions and returns only the next cheapest axiom template according to the dynamic ordering function. Hence, we have $at_i := \mathsf{order}(At^j, R_j)$ instead of $at_1, \ldots, at_n := \mathsf{order}(At^j)$. We further insert a line after calling **order** to remove the cheapest axiom template from the current component: $At^j := At^j \setminus \{at_i\}$. As a result, the next iteration of the for loop will compute the cheapest axiom template amongst the not yet evaluated templates until, in the last iteration, we only have one axiom template left.

Algorithm 3 is sound and complete. The soundness and completeness of the algorithm is based on the following facts:

- The method **rewrite** (see Definition 13) does not affect the answers to a query $q$, since it rewrites axiom templates to templates with the same set of answers.

- The method **connectedComponents** does not affect the answers of $q$; it just splits the query into several components that are evaluated separately and we then take the cartesian product of the answers.

- The method **order** does not change the query in any way; it just reorders the axiom templates.

---

3. We say that a concept name $A$ is a direct subconcept of a concept name $B$ w.r.t. $\mathcal{O}$, if $\mathcal{O} \models A \sqsubseteq B$ and there is no other concept name $A'$ such that $\mathcal{O} \models A' \sqsubseteq B$, $\mathcal{O} \models A \sqsubseteq A'$ and $\mathcal{O} \not\models A \equiv A'$. In a similar way we can define the direct superconcept, the direct subrole and direct superrole.

- For the actual axiom template evaluation, we iterate over all the templates of the query by taking into account the mappings that have already been computed from the evaluation of previous templates and we distinguish between three cases:

  1. The axiom template is a simple one and contains unbound variables. We use specialized reasoner tasks to compute entailed mappings and since we use a sound and complete reasoner the result is indeed sound and complete.

  2. The axiom template does not contain unbound variables. In this case, we simply check entailment using a sound and complete reasoner.

  3. The axiom template is a complex template that has at least one variable unbound. For variables for which the optimization of Section 6.2 is applicable, we initialize the variables to $\top/\top_r$ ($\bot/\bot_r$) and we traverse the concept/role hierarchy top-down (bottom-up). We prune mappings according to Theorems 1 and 2 in case a checked mapping does not constitute a solution mapping. In this case, we do not extend the set of possible mappings $B$. For variables of axiom templates to which the hierarchy optimization is not applicable, we check all compatible mappings. Thus, due to Theorem 1 and 2 the procedure is sound and complete.

Although the above algorithm was implemented in the HermiT reasoner, one can compute the answers of a query using any (hyper)tableau reasoner.

## 7. Evaluation

We tested the developed optimizations with standard benchmarks and a range of custom queries that test complex axiom template evaluation over more expressive ontologies. All experiments were performed on a Mac OS X Lion machine with a 2.53 GHz Intel Core i7 processor and Java 1.6 allowing 1GB of Java heap space. We measure the time for one-off tasks such as classification separately since such tasks are usually performed before the system accepts queries. The ontologies and all code required to perform the experiments are available online (Kollia & Glimm, 2013). The developed system (Glimm & Kollia, 2013), called OWL-BGP, is implemented as a SPARQL Wrapper that can be used with any reasoner that implements the OWLReasoner interface of the OWL API (Horridge & Bechhofer, 2009). In Section 7.1 we compare the different ordering strategies that have been developed on two benchmarks (LUBM and UOBM) that contain queries with variables only in place of individuals (query atoms). We also show the effect of ordering on LUBM using some custom queries with simple axiom templates created for SPARQL-DL (Kremen & Sirin, 2008). In Section 7.2 we show the effect of the proposed optimizations for queries with complex axiom templates. For the evaluation we have used the HermiT hypertableau reasoner (Motik, Shearer, Glimm, Stoilos, & Horrocks, 2013). Other reasoners such as Pellet (Clark & Parsia, 2013a) or Racer Pro (Racer Systems GmbH & Co. KG, 2013) could equally well be used with our implementation as long as they provide an interface with the required statistics, i.e., the number of known and possible instances of concepts and roles for the computation of the cost functions used for query ordering. Without any optimizations, providing this interface with statistics can easily be realized as described in the current paper. The presented query ordering techniques can also be used when optimizations such as caching, pseudo model merging techniques, binary instance retrieval, or absorption are employed. The cost functions might, however, require some adaptation to take the reduction in the required number of consistency checks into account. For example, Pellet uses binary instance retrieval, where testing possible instances of a concept $A$ is realized by splitting the candidate instances into two partitions. For each partition, a single consistency check is performed. If the consistency check is successful, it is safe to consider all individuals belonging to the partition as non-instances of the tested concept $A$. Otherwise, we further split the partition and process the resulting partitions in the same way. In this case, one performs one consistency check to potentially determine several (non-)instances of $A$, which should be reflected in the cost functions.

It is also worth noting that the TrOWL reasoning framework (Thomas, Pan, & Ren, 2013) started to use our SPARQL wrapper to provide SPARQL support. An adaptation to also provide statistics is, to the best of our knowledge, still outstanding, although this should be straightforward. TrOWL is based on two approximate reasoners: one that underapproximates (computation of concept and role instances is sound, but incomplete) (Ren, Pan, & Zhao, 2010) and one that overapproximates (computation of concept and role instances is complete, but unsound) (Pan, Thomas, & Zhao, 2009). In such a setting, the underapproximation can straightforwardly be seen as the known instances and the overapproximation minus the underapproximation as the possible instances.

## 7.1 Query Ordering

We tested our ordering techniques with the Lehigh University Benchmark (LUBM) (Guo, Pan, & Heflin, 2005) as a case where no disjunctive information is present and with the more expressive University Ontology Benchmark (UOBM) (Ma, Yang, Qiu, Xie, Pan, & Liu, 2006).

We first used the 14 conjunctive ABox queries provided in LUBM. From these, queries 2, 7, 8, 9 are the most interesting ones in our setting since they contain many atoms and ordering them can have an effect in running time. We tested the queries on LUBM(1,0) and LUBM(2,0) which contain data for one or two universities respectively, starting from index 0. LUBM(1,0) contains 17,174 individuals and LUBM(2,0) contains 38,334 individuals. LUBM(1,0) took 19 s to load and 0.092 s for classification and initialization of known and possible instances of concepts and roles. The clustering approach for concepts took 1 s and resulted in 16 clusters. The clustering approach for roles lasted 4.9 s and resulted in 17 role successor clusters, 29 role predecessor clusters and 87 role clusters. LUBM(2,0) took 48.5 s to load and 0.136 s for classification and initialization of known and possible instances. The clustering approach for concepts took 3.4 s and resulted in 16 clusters. The clustering approach for roles lasted 16.3 s and resulted in 17 role successor clusters, 31 role predecessor clusters and 102 role clusters. Table 2 shows the execution time for each of the four queries for LUBM(1,0) and LUBM(2,0) for four cases: i) when we use the static algorithm (columns 2 and 6), ii) when we use the dynamic algorithm (columns 3 and 7), iii) when we use random sampling, i.e., taking half of the individuals that are returned (from the evaluation of previous query atoms) in each run, to decide about the next cheapest atom to be evaluated in the dynamic case and iv) using the proposed sampling approach that is based on clusters constructed from individuals in the queried ontology (columns 4 and 8). The queries marked with (*) are the queries where the static and dynamic algorithms result in a different ordering. In Queries 7 and 8 we observe an increase in running time when the dynamic technique is used (in comparison to the static) which is especially evident on Query 8 of LUBM(2,0), where the number of individuals in the ontology and the intermediate result sizes are larger. Dynamic ordering also behaves worse than static in Queries 2 and 9. This happens because, although the dynamic algorithm chooses a better ordering than the static algorithm, the intermediate results (that need to be checked in each iteration to determine the next query atom to be executed) are quite large and hence the cost of iterating over all possible mappings in the dynamic case far outweighs the better ordering that is obtained. We also observe that a random sampling for collecting the ordering statistics in the dynamic case (checking only 50% of individuals in $\Omega_{i-1}$ randomly for detecting the next query atom to be executed) leads to much worse results in most queries than plain static or dynamic ordering. This happens since random sampling often leads to the choice of a worse execution order. The use of the cluster based sampling method performs better than the plain dynamic algorithm in all queries. In Queries 2 and 9, the gain we have from the better ordering of the dynamic algorithm when sampling is used is much more evident. This is the case since we use at most one individual from every cluster for the cost functions computation and the number of clusters is much smaller than the number of the otherwise tested individuals in each run.

| | LUBM(1,0) | | | | LUBM(2,0) | | | |
|---|---|---|---|---|---|---|---|---|
| Q | Static | Dynamic | RSampling | CSampling | Static | Dynamic | RSampling | CSampling |
| *2 | 51 | 119 | 390 | 37 | 162 | 442 | 1,036 | 153 |
| 7 | 25 | 29 | 852 | 20 | 70 | 77 | 2,733 | 64 |
| 8 | 485 | 644 | 639 | 551 | 622 | 866 | 631 | 660 |
| *9 | 1,099 | 2,935 | 3,021 | 769 | 6,108 | 23,202 | 14,362 | 3,018 |

Table 2: Query answering times in milliseconds for LUBM(1,0) and LUBM(2,0) using i) the static algorithm ii) the dynamic algorithm, iii) 50% random sampling (RSampling), iv) the constructed individual clusters for sampling (CSampling)

| Q | PlansNo | Chosen Plan Order | | | Pellet Plan | Worst Plan |
|---|---|---|---|---|---|---|
| | | Static | Dynamic | Sampling | | |
| 2 | 336 | 2 | 1 | 1 | 51 | 4,930 |
| 7 | 14 | 1 | 1 | 1 | 25 | 7,519 |
| 8 | 56 | 1 | 1 | 1 | 495 | 1,782 |
| 9 | 336 | 173 | 160 | 150 | 1,235 | 5,388 |

Table 3: Statistics about the constructed plans and chosen orderings and running times in milliseconds for the orderings chosen by Pellet and for the worst constructed plans

In order to show the effectiveness of our proposed cost functions we compared the running times of all the valid plans (plans that comply to the connectedness condition of Definition 10, i.e., plans on which consecutive atoms share at least one common variable) with the running time of the plan chosen by our method. In the following we show the results for LUBM(1, 0), but the results for LUBM(2,0) are comparable. In Table 3 we show, for each query, the number of valid plans that were constructed according to Definition 10 (column 2), the order of the plan chosen by the static, dynamic, and cluster based sampling methods if we order the valid plans by their execution time (columns 3,4,5; e.g., a value of 2 indicates that the ordering method chose the second best plan), the running time of HermiT for the plan that was created by Pellet (column 6) as well as the running time of the worst constructed plan (column 7).

| LUBM(3,0) | LUBM(4,0) | LUBM(5,0) | LUBM(6,0) | LUBM(7,0) | LUBM(8,0) | LUBM(9,0) |
|-----------|-----------|-----------|-----------|-----------|-----------|-----------|
| 55,664    | 78,579    | 102,368   | 118,500   | 144,612   | 163,552   | 183,425   |

Table 4: Number of individuals in LUBM with increasing number of universities

The comparison of our ordering approach with the approach followed by other reasoners that support conjunctive query answering such as Pellet or Racer Pro is not very straightforward. This is the case because Pellet and Racer have many optimizations for instance retrieval (Sirin et al., 2007; Haarslev & Möller, 2008), which HermiT does not have. Thus, a comparison between the execution times of these reasoners and HermiT would not convey much information about the effectiveness of the proposed query ordering techniques. The idea of comparing only the orderings computed by other reasoners with those computed by our methods is also not very informative since the orderings chosen by different reasoners depend much on the way that queries are evaluated and on the costs of specific tasks in these reasoners and, hence, are reasoner dependent, i.e., an ordering that is good for one reasoner and which leads to an efficient evaluation may not be good for another reasoner. We should note that when we were searching for orderings according to Pellet, we switched off the simplification optimization that Pellet implements regarding the exploitation of domain and range axioms of the queried ontology for reducing the number of query atoms to be evaluated (Sirin & Parsia, 2006). This has been done in order to better evaluate the difference of the plain ordering obtained by Pellet and HermiT since our cost functions take into account all the query atoms.

We observe that for all queries apart from Query 9 the orderings chosen by our algorithms are the (near)optimal ones. For Queries 2 and 7, Pellet chooses the same ordering as our algorithms. For Query 8, Pellet chooses an ordering which, when evaluated with HermiT, results in higher execution time. For Query 9, our algorithms choose plans from about the middle of the order over all the valid plans w.r.t. the query execution time, which means that our algorithms do not perform well in this query. This is because of the greedy techniques we have used to find the execution plan which take into account only local information to choose the next query atom to be executed. Interestingly, the use of cluster based sampling has led to the finding of a better ordering, as we can see from the running time in Table 2 and the better ordering of the plan found with cluster based sampling techniques compared to static or plain dynamic ordering (Table 3). The ordering chosen by Pellet for Query 9 does also not perform well. We see that, in all queries, the worst running times are many orders of magnitude greater than the running times achieved by our ordering algorithms. In general, we observe that in LUBM static techniques are adequate and the use of dynamic ordering does not improve the execution time much compared to static ordering.

| Q | LUBM(3,0) | LUBM(4,0) | LUBM(5,0) | LUBM(6,0) | LUBM(7,0) | LUBM(8,0) | LUBM(9,0) |
|---|-----------|-----------|-----------|-----------|-----------|-----------|-----------|
| 2 | 0.35 | 0.62 | 1.26 | 1.71 | 2.26 | 3.11 | 4.18 |
| 7 | 0.11 | 0.16 | 0.23 | 0.32 | 0.33 | 0.33 | 0.40 |
| 8 | 0.77 | 0.91 | 1.27 | 1.29 | 1.34 | 1.44 | 1.65 |
| 9 | 18.49 | 42.98 | 85.54 | 116.88 | 181.07 | 235.06 | 312.71 |
| all | 20.64 | 55.16 | 90.99 | 138.84 | 213.59 | 241.85 | 323.15 |

Table 5: Query answering times in seconds for LUBM with increasing number of universities

| Q | Static | Dynamic | CSampling | PlansNo | Chosen Plan Order | | | Pellet Plan | Worst Plan |
|---|--------|---------|-----------|---------|-------|---------|----------|-------------|------------|
| | | | | | Static | Dynamic | Sampling | | |
| 4 | 13.35 | 13.40 | 13.41 | 14 | 1 | 1 | 1 | 13.40 | 271.56 |
| 9 | 186.30 | 188.58 | 185.40 | 8 | 1 | 1 | 1 | 636.91 | 636.91 |
| 11 | 0.98 | 0.84 | 1.67 | 30 | 1 | 1 | 1 | 0.98 | > 30 min |
| 12 | 0.01 | 0.01 | 0.01 | 4 | 1 | 1 | 1 | 0.01 | > 30 min |
| 14 | 94.61 | 90.60 | 93.40 | 14 | 2 | 1 | 1 | > 30 min | > 30 min |
| $q_1$ | 191.07 | 98.24 | 100.25 | 6 | 2 | 1 | 1 | > 30 min | > 30 min |
| $q_2$ | 47.04 | 22.20 | 22.51 | 6 | 2 | 1 | 1 | 22.2 | > 30 min |

Table 6: Query answering times in seconds for UOBM (1 university, 3 departments) and statistics

In order to show the scalability of the system, we next run the LUBM queries with different numbers of universities, i.e., LUBM(i,0) where $i$ ranges from 3 to 9. Table 4 shows the number of individuals appearing in each ABox of different university size. The running times of Queries 2, 7, 8, 9 as well as the running time of all the 14 LUBM queries are shown in Table 5. The results for LUBM(1,0) and LUBM(2,0) are shown in Table 2. Note that the results shown are for the case that static ordering is performed. From this table we see that for all queries, the running time increases when the number of individuals of the ABox increases, which is reasonable. We observe that query answering over ontologies is still not as scalable as query answering over databases and this is so, because of the more expressive schema that has to be taken into account and the fact that we have incomplete information in contrast to databases where we have complete information.

Unlike LUBM, the UOBM ontology contains disjunctions and the reasoner makes also nondeterministic derivations. In order to reduce the reasoning time, we removed the nominals and only used the first three departments containing 6,409 individuals. The resulting ontology took 16 s to load and 0.1 s to classify and initialize the known and possible instances. The clustering approach for concepts took 1.6 s and resulted in 356 clusters. The clustering approach for roles lasted 6.3 s and resulted in 451 role successor clusters, 390 role predecessor clusters and 4,270 role clusters. We present results for the static and dynamic algorithms on Queries 4, 9, 11, 12 and 14 provided in UOBM, which are the most interesting ones because they consist of many atoms. Most of these queries contain one atom with possible instances. As we see from Table 6, static and dynamic ordering show similar performance in Queries 4, 9, 11 and 12. Since the available statistics in this case are quite accurate, both methods find the optimal plans and the intermediate result set sizes are small. For both ordering methods, atoms with possible instances for these queries are executed last. In Query 14, the dynamic algorithm finds a better ordering which results in comparable performance. The effect that the cluster based sampling technique has on the running time is not as obvious as in the case of LUBM. This happens because in the current experiment the intermediate result sizes are not very large and, most importantly, because the gain obtained due to sampling is in the order of milliseconds whereas the total query answering times are in the order of seconds obscuring the small improvement in running time due to sampling. In all queries the orderings created by Pellet result in the same or worse running times than the orderings created by our algorithms.

In order to illustrate when dynamic ordering performs better than static, we also created the two custom queries:

$$q_1 = \{ \text{isAdvisedBy}(?x,?y), \text{GraduateStudent}(?x), \text{Woman}(?y) \}$$
$$q_2 = \{ \text{SportsFan}(?x), \text{GraduateStudent}(?x), \text{Woman}(?x) \}$$

In both queries, $P[\text{GraduateStudent}]$, $P[\text{Woman}]$ and $P[\text{isAdvisedBy}]$ are non-empty, i.e., the query concepts and roles have possible instances. The running times for dynamic ordering are smaller since the more accurate statistics result in a smaller number of possible instances that have to be checked during query execution. In particular, for the static ordering, 151 and 41 possible instances have to be checked in query $q_1$ and $q_2$, respectively, compared to only 77 and 23 for the dynamic ordering. Moreover, the intermediate results are generally smaller in dynamic ordering than in static leading to a significant reduction in the running time of the queries. Interestingly, query $q_2$ could not be answered within the time limit of 30 minutes when we transformed the three query concepts into a conjunction, i.e., when we asked for instances of the intersection of the three concepts. This is because for complex concepts the reasoner can no longer use the information about known and possible instances and falls back to a more naive way of computing the concept instances. Again, for the same reasons as before, the sampling techniques have no apparent effect on the running time of these queries.

For each query of the SPARQL-DL tests issued over LUBM(1,0) (Kremen & Sirin, 2008) (cf. Table 7), Table 8 shows the running time of the plan chosen by our method (column 2), the number of valid plans, i.e., plans that comply to the connectedness condition of Definition 10 (column 3), the order of the chosen plan if we order the valid plans by their execution times (column 4), the running time of HermiT for the plan that was created by Pellet (column 5) as well as the running time of the worst constructed plan (column 6). The queries as shown in Table 7 are ordered according to our static ordering algorithm. Since reasoning for LUBM is deterministic, we use static planning to order the axiom templates. Dynamic planning does not improve the execution times (actually it makes them worse) since, as it has been explained before, with only deterministic reasoning we have most of the important information for ordering from the beginning and the overhead caused by dynamic ordering results in worse query execution time.

From the results of Table 8 one can observe that for Queries 1, 2, 3, 4 and 8 the proposed ordering chooses the optimal plan among all valid plans. For Queries 5, 6, 7, 9 and 10 the optimal plan is not chosen according to the proposed cost estimation algorithm. For Queries 5, 7, 9 and 10 this is due to the greedy techniques we have used for finding in

| 1 | GraduateStudent(?x) |
|---|---|
| | ?y(?x, ?z) |
| | Course(?w) |

| 2 | ?c ⊑ Employee |
|---|---|
| | ?c(?x) |
| | Student(?x) |
| | undergraduateDegreeFrom(?x, ?y) |

| 3 | ?y ⊑ memberOf |
|---|---|
| | ?y(?x, University0) |
| | Person(?x) |

| 4 | ?y(GraduateStudent5, ?w) |
|---|---|
| | ?z(?w) |
| | ?z ⊑ Course |

| 5 | ?z ⊑ Course |
|---|---|
| | ?z(?w) |
| | ?y(?x, ?w) |
| | GraduateStudent(?x) |

| 6 | GraduateStudent(?x) |
|---|---|
| | ?y(?x, ?w) |
| | ?z(?w) |
| | GraduateCourse ⊑ ¬?z |

| 7 | ?c ⊑ ⊤ |
|---|---|
| | ?c(?x) |
| | teachingAssistantOf(?x, ?y) |
| | takesCourse(?x, ?y) |

| 8 | ?c ⊑ Person |
|---|---|
| | ?c(?x) |
| | advisor(?x, ?y) |

| 9 | ?c ⊑ Person |
|---|---|
| | ?c(?x) |
| | teachingAssistantOf(?x, ?y) |
| | Course(?y) |

| 10 | ?p ⊑ worksFor |
|---|---|
| | ?p(?y, ?w) |
| | ?c(?y) |
| | ?c ⊑ Faculty |
| | advisor(?x, ?y) |
| | GraduateStudent(?x) |
| | memberOf(?x, ?w) |

Table 7: Queries used for SPARQL-DL tests

| Query | Chosen Ordering Time | PlansNo | Chosen Plan Order | Pellet Plan Time | Worst Plan Time |
|---|---|---|---|---|---|
| 1 | 0.36 | 2 | 1 | 0.36 | 0.58 |
| 2 | 0.03 | 14 | 1 | 0.37 | 0.61 |
| 3 | 0.05 | 4 | 1 | 5.44 | 5.45 |
| 4 | 0.01 | 4 | 1 | 0.01 | 11.46 |
| 5 | 26.10 | 8 | 5 | 0.95 | 454.25 |
| 6 | 10.49 | 8 | 2 | 10.49 | 499.65 |
| 7 | 0.42 | 14 | 6 | 2.68 | 2.68 |
| 8 | 0.23 | 4 | 1 | 0.23 | 0.80 |
| 9 | 0.19 | 8 | 4 | 0.19 | 0.47 |
| 10 | 0.80 | 812 | 21 | 0.80 | 992.77 |

Table 8: Query answering times in seconds for the queries of Table 7 over LUBM(1,0) and statistics

each iteration of our ordering algorithm the next cheapest axiom template to be evaluated. For example, the optimal plan for Query 10 starts with the template GraduateStudent(?x), which is not the cheapest one according to our cost based technique and then, while moving over connected templates, a different order is chosen than the order chosen by our algorithm. It turns out that all valid plans beginning with the atom GraduateStudent($?x$) lead to better execution times than the plan chosen by our algorithm resulting in the existence of several better plans than the chosen one.

For Query 6 we do not find the optimal plan because we have overestimated the cost of the disjoint axiom template and hence have missed the optimal ordering. Nevertheless, the chosen plans lead to execution times for all queries that are up to three orders of magnitude lower than those when the worst plans are chosen. For queries in which the proposed ordering does not lead to the optimal plan, one has to additionally take into account the time we saved from not computing the costs for the $|q!|$ possible orderings, which can be very high. Apart from Queries 4, 6 and 8, we observe that the plans produced by Pellet are not optimal when evaluated with HermiT. As we have discussed before, this happens because the statistics created for ordering are reasoner specific and hence a good ordering for one reasoner may not be good for another reasoner.

## 7.2 Complex Axiom Template Optimizations

In the absence of suitable standard benchmarks for arbitrary SPARQL queries, we created a custom set of queries as shown in Tables 10 and 12 for the GALEN and the FBbt_XP ontology, respectively. Systems that fully support the SPARQL Direct Semantics entailment regime are still under development, which makes it hard to compare our results for these kinds of queries with other systems.

GALEN is a biomedical ontology. It's expressivity is (Horn-)$\mathcal{SHIF}$ and it consists of 2,748 concepts and 413 abstract roles. FBbt_XP is an ontology taken from the Open Biological Ontologies (OBO) Foundry (OBO Foundry, 2013). It falls into the $\mathcal{SHI}$ fragment of $\mathcal{SROIQ}$ and consists of 7,221 concepts and 21 abstract roles. We only consider the TBox part of FBbt_XP since the ABox is not relevant for showing the different effects of the proposed optimizations on the execution times of the considered queries. GALEN took 3.7 s to load and 11.1 s to classify (concepts and roles), while FBbt_XP took 1.5 s to load and 7.4 s to classify.

The execution times for the queries of Tables 10 and 12 are shown on the right-hand side of Tables 9 and 11, respectively. We have set a time limit of 30 minutes for each query. For each query, we tested the execution once without optimizations and once for each combination of applicable optimizations from Sections 5 and 6. In Tables 9 and 11, one can also see the number of consistency checks that were performed for the evaluation of each query and each combination of the applicable optimizations as well as the number of results of each query. In these tables we have taken the time of the worst ordering of query atoms for the cases in which the ordering optimization is applicable but not enabled. Note that only the complex axiom templates require consistency checks to be evaluated; the simple ones (subsumption axiom templates in this case) need only cache lookups in the reasoner's internal structures since the concepts and roles are already classified.

**GALEN Queries:** As expected, an increase in the number of variables within an axiom template leads to a significant increase in the query execution time because the number of mappings that have to be checked grows exponentially in the number of variables. This can, in particular, be observed from the difference in execution time between Query 1 and 2.

| Query | Reordering | Hierarchy Exploitation | Rewriting | Consistency Checks | Time | AnswersNo |
|---|---|---|---|---|---|---|
| 1 | | | | 2,750 | 1.68 | 10 |
| 1 | | x | | 50 | 0.18 | 10 |
| 2 | | | | 1,141,250 | 578.98 | 214 |
| 2 | | x | | 1,291 | 9.85 | 214 |
| 3 | | x | | | >30 min | |
| 3 | | | x | 19,250 | 102.37 | 2,816 |
| 3 | | x | x | 3,073 | 2.69 | 2,816 |
| 4 | x | x | | | > 30 min | |
| 4 | | x | x | | > 30 min | |
| 4 | x | | x | 16,135 | 7.68 | 51 |
| 4 | x | x | x | 197 | 1.12 | 51 |
| 5 | | x | | | >30 min | |
| 5 | x | | | 1,883 | 0.67 | 4,392 |
| 5 | x | x | | 1,883 | 0.8 | 4,392 |

Table 9: Query answering times in seconds for the queries of Table 10 with and without optimizations

| | |
|---|---|
| 1 | $Infection \sqsubseteq \exists hasCausalLinkTo.?x$ |
| 2 | $Infection \sqsubseteq \exists ?y.?x$ |
| 3 | $?x \sqsubseteq Infection \sqcap \exists hasCausalAgent.?y$ |
| 4 | $NAMEDLigament \sqsubseteq NAMEDInternalBodyPart \sqcap ?x$ |
| | $?x \sqsubseteq \exists hasShapeAnalagousTo?y \sqcap \exists ?z.linear$ |
| 5 | $?x \sqsubseteq NonNormalCondition$ |
| | $?z \sqsubseteq ModifierAttribute$ |
| | $Bacterium \sqsubseteq \exists ?z.?w$ |
| | $?y \sqsubseteq StatusAttribute$ |
| | $?w \sqsubseteq AbstractStatus$ |
| | $?x \sqsubseteq \exists ?y.Status$ |

Table 10: Sample complex queries for the GALEN ontology

From these two queries, it is evident that the use of the hierarchy exploitation optimization leads to a decrease in execution time of up to two orders of magnitude. Query 3 can only be completed in the time limit if at least the query rewriting optimization is enabled. We can get an improvement of up to three orders of magnitude in this query, by using rewriting in combination with the hierarchy exploitation. Query 4 can only be completed in the given time limit if at least reordering and rewriting is enabled. Rewriting splits the first axiom template into the following two simple axiom templates, which are evaluated much more efficiently:

$$NAMEDLigament \sqsubseteq NAMEDInternalBodyPart \qquad and \qquad NAMEDLigament \sqsubseteq ?x$$

After the rewriting, the ordering optimization has an even more pronounced effect since both rewritten axiom templates can be evaluated with a simple cache lookup. Without ordering, the complex axiom template could be executed before the simple ones, which leads to the inability of answering the query within the time limit of 30 min. Without a good ordering, Query 5 can also not be answered within the time limit. The ordering chosen by our algorithm is shown below. Note that the query consists of two connected components: one for the axioms containing $?z$ and $?w$ and another one for the axioms containing $?x$ and $?y$.

$$?z \sqsubseteq \text{ModifierAttribute}$$
$$?w \sqsubseteq \text{AbstractStatus}$$
$$\text{Bacterium} \sqsubseteq \exists?z.?w$$

$$?y \sqsubseteq \text{StatusAttribute}$$
$$?x \sqsubseteq \text{NonNormalCondition}$$
$$?x \sqsubseteq \exists?y.\text{Status}$$

In this query, the hierarchy exploitation optimization does not improve the execution time since, due to the chosen ordering, the variables on which the hierarchy optimization can be applied, are already bound when it comes to the evaluation of the complex templates. Hence, the running times with and without the hierarchy exploitation are similar. The number of consistency checks is significantly lower than the number of answers because the overall results are computed by taking the cartesian products of the results for the two connected components. Interestingly, for queries with complex axiom templates, it does not make sense to require that the next axiom template to evaluate shares a variable with the previously evaluated axiom templates, as in the case of simple axiom templates. For example, if we would require that, the first connected component of the query would be executed in the following order:

$$?z \sqsubseteq \text{ModifierAttribute}$$
$$\text{Bacterium} \sqsubseteq \exists?z.?w$$
$$?w \sqsubseteq \text{AbstractStatus}$$

this results in 294,250 instead of 1,498 consistency checks since we no longer use a cheap cache look-up check to determine the bindings for $?w$, but first iterate over all possible $?w$ bindings and check entailment of the complex axiom template and then reduce the computed candidates when processing the last axiom template.

Although our optimizations can significantly improve the query execution time, the required time can still be quite high. In practice, it is, therefore, advisable to add as many restrictive axiom templates (axiom templates which require only cache lookups) for query variables as possible. For example, the addition of $?y \sqsubseteq$ Shape to Query 4 reduces the runtime from 1.12 s to 0.65 s. We observe, as expected, that the execution time for each query and applicable optimization is analogous to the number of consistency checks that are performed for the evaluation of the query.

**FBbt_XP Queries:** For Queries 1, 2, 3, 5 and 6, on which the ordering optimization is applicable, we observe a decrease in execution time up to two orders of magnitude when the ordering optimization is used. The ordering optimization is important for answering Queries 1, 2 and 3 within the time limit. For all queries, the additional use of the hierarchy exploitation optimization leads to an improvement of up to three orders of magnitude. We observe that in some queries the effect of the hierarchy exploitation is more profound than in others. More precisely, the smaller the ratio of the result size to the number of consistency checks without the hierarchy optimization, the more pronounced is the effect when enabling this optimization. In other words, when more tested mappings are indeed solutions, one can prune fewer parts of the hierarchy since pruning can only be performed when we find a non-solution. In Query 2, we even observe a slight increase in running time when the hierarchy optimization is used. This is because the optimization can only prune few candidate mappings, which does not outweigh the overhead caused by maintaining information about which hierarchy parts have already been tested. In Query 6, the rewriting optimization is important to answer the query within the time limit. When all optimizations are enabled, the number of consistency checks is less than the result size (32,490 versus 43,338) since only the complex axiom template requires consistency checks.

| Query | Reordering | Hierarchy Exploitation | Rewriting | Consistency Checks | Time | AnswersNo |
|---|---|---|---|---|---|---|
| 1 | x | | | 151,683 | 44.13 | 7,243 |
| 1 | | x | | | > 30 min | |
| 1 | x | x | | 11,262 | 5.64 | 7,243 |
| 2 | x | | | 14,446 | 37.38 | 7,224 |
| 2 | | x | | | > 30 min | |
| 2 | x | x | | 12,637 | 39.20 | 7,224 |
| 3 | x | | | 72,230 | 357.59 | 188 |
| 3 | | x | | | > 30 min | |
| 3 | x | x | | 54,186 | 252.41 | 188 |
| 4 | | | | 166,129 | 486.81 | 68 |
| 4 | | x | | 1335 | 17.03 | 68 |
| 5 | | | | 166,129 | 457.84 | 0 |
| 5 | x | | | 21,669 | 19.68 | 0 |
| 5 | | x | | 907 | 11.74 | 0 |
| 5 | x | x | | 3 | 0.01 | 0 |
| 6 | x | x | | | > 30 min | |
| 6 | x | | x | 43,338 | 183.66 | 43,338 |
| 6 | | x | x | | > 30 min | |
| 6 | x | x | x | 32,490 | 152.38 | 43,338 |

Table 11: Query answering times in seconds for the queries of Table 12 with and without optimizations

| | | | |
|---|---|---|---|
| 1 | $?x \sqsubseteq \forall \text{part\_of}.?y$ <br> $?x \sqsubseteq \text{FBbt\_00005789}$ | 4 | $\text{FBbt\_00001606} \sqsubseteq \exists ?y.?x$ |
| 2 | $?y \sqsubseteq \text{part\_of}$ <br> $?x \sqsubseteq \forall ?y.\text{FBbt\_00001606}$ | 5 | $\text{FBbt\_00001606} \sqsubseteq \exists ?y.?x$ <br> $?y \sqsubseteq \text{develops\_from}$ |
| 3 | $?x \sqsubseteq \exists ?y.\text{FBbt\_00025990}$ <br> $?y \sqsubseteq \text{overlaps}$ | 6 | $?y \sqsubseteq \text{FBbt\_00001884}$ <br> $?p \sqsubseteq \text{part\_of}$ <br> $?x \sqsubseteq \exists ?p.?y \sqcap ?w$ |

Table 12: Sample complex queries for the FBbt_XP ontology

## 8. Related Work

There is not yet a standardized and commonly implemented query language for OWL ontologies. Several of the widely deployed systems support, however, some query language. Pellet supports SPARQL-DL (Sirin & Parsia, 2007), which is a subset of SPARQL, adapted to work with OWL's Direct Semantics. The kinds of SPARQL queries that are supported in SPARQL-DL are those that can directly be mapped to reasoner tasks. Therefore, SPARQL-DL can be understood as queries that only use simple axiom templates in our terminology. Similarly, KAON2 (Hustadt, Motik, & Sattler, 2004) supports SPARQL queries, but restricted to ABox queries/conjunctive instance queries. To the best of our knowledge, there are no publications that describe any ordering strategies for KAON2. Racer Pro (Haarslev & Möller, 2001) has a proprietary query language, called nRQL (Haarslev et al., 2004), which allows for queries that go beyond ABox queries, e.g., one can retrieve sub- or super-concepts of a given concept. TrOWL (Thomas et al., 2013) is another system that supports SPARQL queries, but the reasoning in TrOWL is approximate, i.e., an OWL DL ontology is rewritten into an ontology that uses a less expressive language before reasoning is applied (Thomas, Pan, & Ren, 2010). TrOWL is based on the SPARQL framework presented here, but instead of using HermiT as background reasoner, it uses its approximate reasoners for the OWL 2 EL and OWL 2 QL profiles. Furthermore, there are systems such as QuOnto (Acciarri, Calvanese, De Giacomo, Lembo, Lenzerini, Palmieri, & Rosati, 2013) or Requiem (Pérez-Urbina, Motik, & Horrocks, 2013), which support profiles of OWL 2, and which support conjunctive queries, e.g., written in SPARQL syntax, but with proper non-distinguished variables. Of the systems that support all of OWL 2 DL, only Pellet supports non-distinguished variables as long as they are not used in cycles, since decidability of cyclic conjunctive queries is to the best of our knowledge still an open problem.

The problem of finding good orderings for the templates of a query issued over an ontology has already been preliminarily studied (Sirin & Parsia, 2006; Kremen & Sirin, 2008; Haarslev & Möller, 2008). Similarly to our work, Sirin and Parsia as well as Kremen and Sirin exploit reasoning techniques and information provided by reasoner models to create statistics about the cost and the result size of axiom template evaluations within execution plans. A difference is that they use cached models for cheaply finding obvious concept and role (non-)instances, whereas in our case we do not cache any model or model parts. Instead we process the pre-model constructed for the initial ontology consistency check and extract the known and possible instances of concepts and roles from it. We subsequently use this information to create and update the query atom statistics. Moreover, Sirin and Parsia and Kremen and Sirin compare the costs of complete execution plans —after heuristically reducing the huge number of possible complete plans — and choose the one that is most promising before the beginning of query execution. This is different from our cheap greedy algorithm that finds, at each iteration, the next most promising axiom template. Our experimental study shows that this is equally effective as the investigation of all possible execution orders. Moreover, in our work we have additionally used dynamic ordering combined with clustering techniques, apart from static ones, and have shown that these techniques lead to better performance particularly in ontologies that contain disjunctions and do now allow for purely deterministic reasoning.

Haarslev and Möller discuss by means of an example the ordering criteria they use to find efficient query execution plans in Racer Pro. In particular, they use traditional database cost based optimization techniques, which means that they take into account only the cardinality of concept and role atoms to decide about the most promising ordering. As previously discussed, this can be inadequate especially for ontologies with disjunctive information.

A significant amount of work on the estimation of cost metrics and the search for optimal orders for evaluating joins has been performed in the context of databases. As discussed in Section 3, in databases, cost formulas are defined that estimate the CPU and I/O costs (similar to our reasoning costs) and the number of returned tuples (similar to our result sizes). These estimates are used to find good join orders. The System R query optimizer, for example, is among the first works to use extended statistics and a novel dynamic programming algorithm to find effective (minimal) join orders of query atoms (Selinger, Astrahan, Chamberlin, Lorie, & Price, 1979). A heuristic similar to ours (for the case of conjunctive instance queries) is used in this work, according to which the join order permutations are reduced by avoiding Cartesian products of result sets of query atoms. Regarding join order selection, apart from dynamic programming, also other algorithmic paradigms based on branch-and-bound or simulated annealing have, since then, been presented in the literature. Dynamic ordering has also been explored in the literature in the context of adaptive query processing techniques (Gounaris, Paton, Fernandes, & Sakellariou, 2002), which have been proposed to overcome the problems caused by the lack of necessary statistics, good selectivity estimates, knowledge for the runtime mappings of a query at compile time. These techniques take into account changes that happen to the evaluation environment at runtime and modify the execution plan at runtime (i.e., they change the used operators for joins or the order in which the (remaining) query atoms are evaluated).

## 9. Conclusions

In the current paper, we presented a sound and complete query answering algorithm and novel optimizations for the OWL Direct Semantics entailment regime of SPARQL 1.1. Our prototypical query answering system combines existing tools such as ARQ, the OWL API, and the HermiT OWL reasoner. Apart from the query ordering optimization—which uses (reasoner dependent) statistics provided by HermiT—the system is independent of the reasoner used, and could employ any reasoner that supports the OWL API.

We propose two cost-based ordering strategies for finding (near-)optimal execution orders for conjunctive instance queries. The cost formulas are based on information extracted from models of a reasoner (in our case HermiT). We show through an experimental study that static techniques are quite adequate for ontologies in which reasoning is deterministic. When reasoning is nondeterministic, however, dynamic techniques often perform better.

The use of cluster based sampling techniques can improve the performance of the dynamic algorithm when the intermediate result sizes of queries are sufficiently large, whereas random sampling is not beneficial and often leads to suboptimal query execution plans.

The presented approach can be used to find answers to queries issued over $\mathcal{SROIQ}$ ontologies. Since it is based on entailment checking for finding answers to conjunctive instance queries it is not as scalable as other techniques, such as query rewriting, which are applied to ontologies of lower expressivity, such as DL-Lite. In other words, there is a trade-off between scalability and ontology expressivity and one needs to consider if it is more important for one's application to use a more scalable query answering system with a less expressive ontology or a less scalable system with a more expressive ontology.

The module for ordering is based on the extraction of statistics from a reasoner model, which is computed off-line. Any update of the ontology ABox would then cause the construction of a new model from scratch and the consequent recompilation of known and possible instances of concepts and roles unless an incremental reasoner is used. An incremental reasoner could, for example, find modules of the pre-model that are affected by the update and recompute only model parts. One could then also incrementally update the statistics that are used for ordering. To the best of our knowledge, OWL DL reasoners only partially support incremental reasoning and we have not considered this case in the current paper.

For queries that go beyond conjunctive instance queries we further provide optimizations such as rewriting into equivalent, but simpler queries. Another highly effective and frequently applicable optimization prunes the number of candidate solutions that have to be checked by exploiting the concept and role hierarchies. One can, usually, assume that these hierarchies are computed before a system accepts queries. Our empirical evaluation shows that this optimization can reduce the query evaluation times up to three orders of magnitude.

## Acknowledgments

This work was done within the Transregional Collaborative Research Centre SFB/TRR 62 "A Companion-Technology for Cognitive Technical Systems" funded by the German Research Foundation (DFG).